\documentclass[12pt]{spieman}  
\usepackage{amsmath,amsfonts,amssymb}
\usepackage{graphicx}
\usepackage{setspace}
\usepackage{tocloft}
\usepackage{wrapfig}
\newcommand{\astfootnote}[1]{%
\let\oldthefootnote=\thefootnote%
\setcounter{footnote}{0}%
\renewcommand{\thefootnote}{\fnsymbol{footnote}}%
\footnote{#1}%
\let\thefootnote=\oldthefootnote%
}
\newcommand{\elsa}{\color{black}}
\newcommand{\elsaa}{\color{black}}
\newcommand{\elsaaa}{\color{black}}
\usepackage{lineno}

\title{Photonic beam-combiner for visible interferometry with SCExAO/FIRST: laboratory characterization and design optimization}

\author[a,*]{Manon\,Lallement}
\author[a]{Elsa\,Huby}
\author[a]{Sylvestre\,Lacour}
\author[b]{Guillermo\,Martin}
\author[a]{Kevin\,Barjot}
\author[a]{Guy\,Perrin}
\author[a]{Daniel\,Rouan}
\author[a]{Vincent\,Lapeyrere}
\author[c,d]{Sebastien\,Vievard}
\author[c,e]{Olivier\,Guyon}
\author[c]{Julien\,Lozi}
\author[c]{Vincent\,Deo}
\author[c,d]{Takayuki\,Kotani}
\author[f]{Cecil\,Pham}
\author[f]{Cedric\,Cassagnettes}
\author[f]{Adrien\,Billat}
\author[g]{Nick\,Cvetojevic}
\author[h]{Franck\,Marchis}

\affil[a]{LESIA, Observatoire de Paris, Université PSL, CNRS, Sorbonne Université, Université Paris Cité, 5 place Jules Janssen, 92195 Meudon, France}
\affil[b]{Univ. Grenoble Alpes, CNRS, IPAG, 38000 Grenoble, France}
\affil[c]{National Astronomical Observatory of Japan (NAOJ), 650 North A'ohoku Place, Hilo, Hawaii 96720, United States}
\affil[d]{Astrobiology Center, 2-21-1, Osawa, Mitaka, Tokyo, 181-8588, Japan}
\affil[e]{University of Arizona, Tucson, AZ 85721, United States}
\affil[f]{TEEM Photonics, 61 Chem. du Vieux Chêne, Meylan, France}
\affil[g]{Côte d'Azur Observatory, 96 Bd de l'Observatoire, 06300 Nice, France}
\affil[h]{Search for Extra-Terrestrial Intelligence (SETI), 339 Bernardo Ave, Mountain View, CA 94043, United States}

\cftpagenumbersoff{figure}
\cftpagenumbersoff{table} 
\begin{document} 
\maketitle

\begin{abstract}
Integrated optics are used to achieve astronomical interferometry inside robust and compact materials, improving the instrument's stability and sensitivity. In order to perform differential phase measurements at the H$\alpha$ line (656.3\,nm) with the 600-800\,nm spectro-interferometer FIRST, a photonic integrated circuit (PIC) is being developed {\color{black}{in collaboration with TEEM Photonics.}} This PIC performs the {\elsa{interferometric}} combination of the beams coming from sub-apertures {\elsa{selected in}} the telescope pupil, {\elsa{thus implementing the pupil remapping technique to restore the diffraction limit of the telescope.}} In this work, {\elsaaa{we report on the latest developments {\elsaaa{carried out within the FIRST project}} to produce a high performance visible PIC. The PICs are manufactured by}} TEEM Photonics, using {\elsaaa{their technology based on}} $K_+:Na_+$ ion exchange in glass. {\elsaaa{The first part of the study consists in the experimental characterization of the fundamental properties of the waveguides, in order to build an accurate model, which is the basis for the design of more complex functions. In the second part, theoretical designs and their optimization for three types of combiner architectures are presented: symmetric directional coupler, asymmetric directional couplers and ABCD cells including achromatic phase shifters.}}
\end{abstract}


{\noindent \footnotesize\textbf{*}Manon Lallement,  \linkable{manon.lallement@obspm.fr}}

\begin{spacing}{2}   

\section{Introduction}
\label{sec:sec1} 
{\color{black}{Interferometric techniques in astronomy}} consist in recovering the source spatial intensity distribution with an angular resolution as fine as $\lambda/2B$, with {\elsa{$B$ the baseline length. Implemented on a monolithic telescope, sparse aperture masking (SAM) is a technique providing a spatial resolution down to $\lambda/2D$ with $\lambda/D$ the telescope diffraction limit. This technique consists in applying a non-redundant mask with holes at the pupil plane defining several sub-apertures}}. At the telescope focal plane, the image no longer corresponds to {\color{black}{the point spread function of the telescope's pupil}} but to the superimposition of fringe patterns created by each pair of interfering {\color{black}{sub-aperture}} beams. The non-redundant configuration of the mask means that each baseline, i.e each pair of {\color{black}{sub-apertures}} separated by a baseline vector $ \vec{B} $, produces a unique fringe pattern. The information carried by each baseline can be retrieved independently. This is well illustrated in the Fourier domain, in which each baseline information, i.e phase and contrast of the associated fringes, is carried by a single peak isolated from the others. All this ensures that there is no blurring effect between the fringes in the presence of {\elsa{residual atmospheric}} turbulence {\elsa{or phase perturbations due to the optical bench}}, and that one can recover information at the diffraction limit of the telescope, and even below. {\color{black}{The SAM technique limitations are (1) the reduced collecting area and thus the sensitivity limit and (2) the speckle noise which remains at the scale of one sub-aperture, limiting the contrast of the fringes}}.

{\color{black}{To address these limitations, the pupil remapping technique\cite{perrin2006high} theoretically gives access to the whole pupil: instead of using a sparse aperture mask, a micro-lens array samples the {\color{black}{whole pupil}} in several sub-apertures and injects their light in single-mode fibers spatially filtering the wavefront, thus removing the speckle noise.}} Sub-aperture pairs are recombined non redundantly or pairwise so that the information carried by each baseline can be retrieved independently. {\elsa{In addition, the interferograms can be spectrally dispersed.}} The spatial intensity distribution of the source is thus recovered with an angular resolution of $\lambda/2D$ {\elsa{with an increased accuracy compared with SAM and with a better (u,v) plane and spectral coverage.}}
FIRST, which stands for Fibered Imager FoR a Single Telescope, was built {\color{black}{at the}} Observatoire de Paris to validate the concept of pupil remapping coupled with spectroscopy. From 2010 to 2013, the instrument was used on the 3m-Shane {\color{black}{Telescope}} at the Lick Observatory{\cite{hubyphd}}. In 2013, FIRST was installed at the 8.2\,$m$ Subaru {\color{black}{Telescope}} \cite{SPIE_2021-Vievard} on the Subaru Coronagraphic Extreme Adaptive Optics platform (SCExAO)\cite{jovanovic2015}, {\elsa{enhancing the ultimate angular resolution of the instrument, with $\lambda/D = 16.5$\,mas at 656.3\,nm.}}
{\elsa{SCExAO delivers a Strehl ratio of 50\% to 60\% in the visible at 750\,nm. FIRST is thus leveraging the wavefront stability provided by SCExAO, making long exposure up to 1 second possible, without loosing the fringe contrast.}}
{\color{black}{Two versions of the FIRST instrument are currently on the SCExAO's bench: FIRST version 1 (FIRSTv1) and FIRST version 2 (FIRSTv2) depending on the interferometric combination method.}} {\elsa{Table~\ref{tab:table2} shows their respective features.}} {\elsa{Close binary stars were detected and spectrally characterized with FIRSTv1 using closure phase measurements \cite{huby2013first,Vievard2023}}}. {\elsa{Currently, our efforts are focusing on pushing the detection limits of the instrument, with the ultimate goal of detecting exoplanetary systems. For development purposes, a replica of FIRSTv2 has been built in the laboratory of Observatoire de Paris\cite{barjot2021}, where the PIC prototypes are characterized prior to further validation on the sky.}} 


\begin{table}[h!]
\caption{{\elsa{Features}} of FIRSTv1 and FIRSTv2 (expected) on SCExAO} 
\label{tab:table2}
\begin{center}       
\begin{tabular}{|l|c|c|}
\hline
\rule[-1ex]{0pt}{3.5ex}  & FIRSTv1 & FIRSTv2 \\
\hline
\rule[-1ex]{0pt}{3.5ex} Spectral band &  \multicolumn{2}{|c|}{600 to 800\,nm} \\
\hline
\rule[-1ex]{0pt}{3.5ex} Number of sub-apertures used & 2 sets of 9 & 5 {\elsa{(goal: $>$ 9)}} \\
\hline
\rule[-1ex]{0pt}{3.5ex} Field of view &  \multicolumn{2}{| c |}{$\sim$ 100\,mas}\\
\hline
\rule[-1ex]{0pt}{3.5ex} Angular resolution $\lambda/2D$ & \multicolumn{2}{| c |}{$\sim$ 8.25\,mas}\\
\hline          
\rule[-1ex]{0pt}{3.5ex} Spectral resolution & 400 & 3600\\
\hline
\rule[-1ex]{0pt}{3.5ex} {\color{black}{Dynamic range\cite{Vievard2023}}} & $10^2$ & {\elsa{goal}:} $10^3$ \\
\hline            
\rule[-1ex]{0pt}{3.5ex} On-sky magnitude limit {\color{black}{in the R band}} & 2.5\,mag & {\elsa{goal: 7\,mag (AB Aurigae)}} \\
\rule[-1ex]{0pt}{3.5ex} {\elsaaa{(to reach the above dynamic range in 1 hour)}} & & \\
\hline
\rule[-1ex]{0pt}{3.5ex} {\elsaaa{Data analysis strategy}} & Closure phase & H$\alpha$ differential phase \\
\hline
\end{tabular}
\end{center}
\end{table}

Young gas giant exoplanets are particularly interesting for FIRSTv2. {\elsa{Studies based on the populations of exoplanets detected by radial velocities\cite{fernandes2019} show that}} the distribution of gas giant is supposed to be maximal for systems with a separation of 1-3\,au, which corresponds to an angular separation {\color{black}{of 7-28\,mas}} at 140\,pc, the distance {\color{black}{to}} the Taurus Nuclear Cloud. {\elsa{This region cannot be probed by classical imager on 8m-telescopes, but the angular resolution offered by the interferometric technique places it within reach of FIRST.}}
Moreover, {\elsa{in the visible, a dynamic range of}} $10^{6}$ to $10^{9}$ is required to differentiate {\color{black}{between the light from an exoplanet and its host star}}\cite{seager2010}. This is currently out of reach for {\elsa{interferometric techniques like}} FIRST. However, at the protoplanetary state, gas giants are {\color{black}{less than 4\,Myr old and are}} still accreting matter from {\elsa{their surrounding disk, inducing a strong emission in the hydrogen line at $656.3$\,nm (H$\alpha$)\cite{aoyama2018,aoyama2019}. As a consequence, the contrast between the planet and the star at this particular line is lowered down to $10^{2}$\,-\,$10^{3}$, making them easier to detect in the visible}}. Three protoplanet detections have been confirmed using H$\alpha$ imaging. This is the case for protoplanets PDS70b and PDS70c detected using the MUSE integral field spectrograph\cite{wagner2018,haffert2019} as well as AB Aurigae b, detected at various wavelengths by several instruments and in particular, at the H$\alpha$ line with VAMPIRES installed on SCExAO\cite{currie2022}. 

{\elsa{With a spectro-interferometer, H$\alpha$ differential phase measurement is the equivalent of H$\alpha$ imaging}}. It consists in comparing the {\color{black}{complex visibility phase}} at the H$\alpha$ line (where the protoplanet is {\elsa{brighter}}) {\color{black}{to the complex visibility phase in the continuum}} (where it is too faint to be detected). This technique {\color{black}{has recently been}} implemented using the high precision phase measurements with the GRAVITY instrument to detect the broad line region around a quasar\cite{Gravity2018}. The FIRST instrument can perform this measurement at high angular resolution {\elsa{in the visible, targeting the H$\alpha$ line. For that purpose, its spectral resolution, sensitivity and {\color{black}{dynamic range}} are currently being enhanced}} {\color{black}{as specified in Table~\ref{tab:table2}.}}
In the FIRST instrument, the telescope's pupil is divided into {\color{black}{sub-apertures}} thanks to a micro-lens array {\color{black}{which couples}} the {\color{black}{light from the sub-apertures}} into {\color{black}{Polarization Maintaining (PM)}} Single-Mode Fibers (SMF). 
Optical delay lines are used to compensate for the fiber length difference and reach the {\color{black}{null}} Optical Path Difference (OPD). In FIRSTv1, SMFs are used to remap the input pupil sub-apertures into a non-redundant configuration in the output pupil\cite{perrin2006high,LacourPhd}. {\elsa{Thanks to this multiplexing}}, each pair of {\elsa{sub-apertures produces a fringe pattern with a unique spatial frequency.}} The interferometric combination of the beams is achieved in free-space following a Young slit-like experiment. FIRSTv2 is currently under development on a testbed at the Observatoire de Paris. In this upgraded version of the instrument, SMF inject the sub-apertures light into a Photonic Integrated Circuit (PIC) where the beams are recombined pairwise. 
The output signal is dispersed with a spectrograph with R $\sim$ 400 in FIRSTv1 and R $\sim$ 3600 in FIRSTv2. 

\begin{figure}
\begin{center}
\begin{tabular}{c}
\includegraphics[height=14cm]{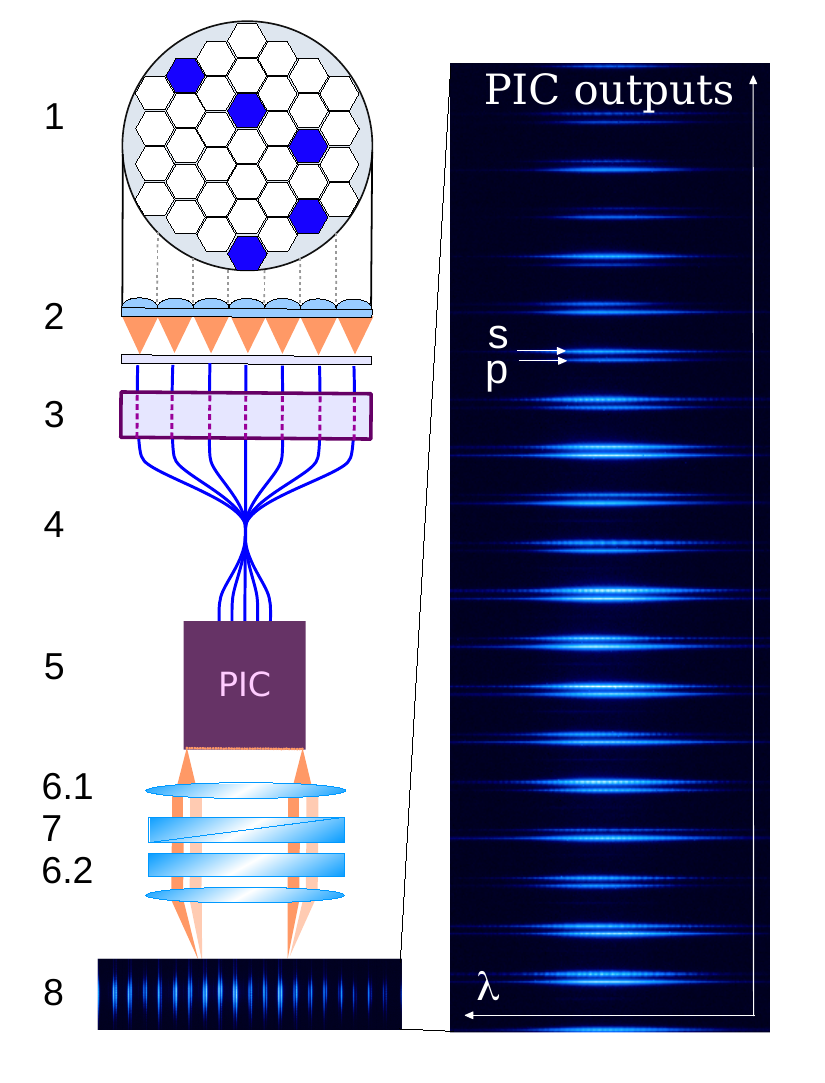}
\end{tabular}
\end{center}
\caption 
{\label{fig:FIRSTv2OpticalSystem} FIRSTv2 optical system. 1: segmented entrance pupil, {\elsa{with the 5 used sub-apertures highlighted in blue}} 2:  micro-lens array injecting the sub-aperture's light in single-mode fibers. 3: optical delay lines (ODL) compensating for the optical path difference between the fibers. 4: single-mode fibers spatially filtering the wavefront and {\color{black}{injecting the sub-apertures light into the photonic integrated circuit (PIC)}}. 5: beam combiner PIC. {\color{black}{6.1: collimating lens of the VPH grating based spectrograph cross-dispersing the PIC outputs light in the horizontal direction. 6.2: VPH grating and imaging lens of the spectrograph.}} 7: Wollaston prism, {\color{black}{consisting of two orthogonal prisms of birefringent material}}, splitting orthogonal polarization{\color{black}{s}} in the vertical direction to avoid fringe blurring. Polarization is {\color{black}{denoted p-polarized (resp. s-polarized) light if polarized along (resp. orthogonally to) the first Wollaston prism's axis. Polarizations}} are represented by dark and light orange beams. 8: spectra obtained on the Zyla sCMOS camera for a given optical path length difference (OPD) far from the null OPD.}
\end{figure}

Fig.~\ref{fig:FIRSTv2OpticalSystem} presents {\color{black}{FIRSTv2 optical system and the sample of data acquired in the laboratory}}: 5 {\color{black}{sub-apertures}} are recombined {\color{black}{accounting for 10 baselines. In the configuration presented in Fig.~\ref{fig:FIRSTv2OpticalSystem}, each baseline is encoded in four horizontal spectra:}} {\color{black}{the interferometric combinations are performed in the PIC by 2x2 directional couplers, i.e there are two outputs per baseline. The light from both PIC outputs is vertically split by a Wollaston prism to avoid fringe blurring {\elsa{induced by the birefringence in the setup}}, and horizontally dispersed on the camera thanks to a Volume Holographic Grating (VPH). It covers the  $600$-$800$\,nm spectral band with a resolution of $R\sim 3600$ at $670$\,nm.}} To further sample the fringes, {\elsa{a segmented mirror is used to temporally modulate the phase of the input beams, by applying piston commands to the segments corresponding to each sub-aperture.}} 

\pagebreak

{\elsa{The PIC device is the heart of the upgraded version of the instrument FIRSTv2, that will enhance the stability, accuracy and sensitivity compared to FIRSTv1. The design and characterization of a high performance visible PIC is a critical step in these developments, and we report in this paper on the latest developments carried out in collaboration with TEEM Photonics to optimize the building blocks required to produce a complete device. Specifications for the PIC component suited for FIRSTv2 are listed in Section~\ref{sec:firstv2},}} and the manufacturing process developed by TEEM Photonics based on $K_+:Na_+$ ion exchange technology (ioNext) is described. In Section~\ref{sec:3model}, the characterization of TEEM Photonics' standard straight waveguides in terms of single-mode spectral range and mode field profile are presented. These measurements are {\elsaaa{critical}} to define a model of the waveguide 3D diffused index profile thanks to the BeamPROP modeling software{\elsaaa{, which is the working basis for the design of more complex functions}}. In Section~\ref{sec:design}, we present {\elsaaa{the theoretical models and design optimization of three types of combiners investigated so far: 1) directional couplers are the simplest design and were characterized in laboratory, showing a weak chromatic behavior, 2) asymmetric directional couplers can be tuned to make the coupler even less chromatic, 3) ABCD cells are more complex and comprise splitters, phase shifters and combiners, but are more convenient as they provide additional measurements of the interference state, thus avoiding the need for temporal modulation of the phase. The latest PIC prototypes and test chips including single combiner functions for characterization and validation are also presented.}} {\color{black}{In Section ~\ref{sec:ccl}, we conclude and present the next steps.}}
\section{{\color{black}{FIRSTv2 visible photonics integrated circuit for high performance beam combination}}}
\label{sec:firstv2} 
 The FIRSTv2's high throughput visible beam combiner is being developed and characterized\cite{martin2016,martin2018} to enhance the stability and {\color{black}{accuracy}} of the measurements of interferometric observables. Compared to {\color{black}{bulk optics}} beam combination, {\color{black}{photonic integrated circuit}} combination is more robust, less sensitive to thermal variations, mechanical constraints and alignment errors. {\color{black}{This is critical to meet FIRSTv2 dynamic range specification presented in Table.~\ref{tab:table2} as phase measurement accuracy directly limits the achievable dynamic range. In FIRSTv1, the free-space interferometric combination leads to spatially sampled fringes\cite{huby2012first}. In FIRSTv2, the fringes are temporally modulated by applying piston commands to a segmented mirror located between the pupil and the micro-lens array. {\elsa{The signal of interest is thus condensed into a few pixels, instead of a few hundreds, reducing the read-out noise and enhancing the phase measurement SNR. Our efforts currently focus on performing the combination of 5 input beams, but for future FIRST upgrades, PIC devices can relatively easily be scaled to a higher number of sub-aperture pairs by densifying the design or by duplicating the PIC devices.}}}}

\subsection{{\elsa{Architecture of the FIRSTv2 photonic beam combiner}}}
\label{subsec:architecture} 
\begin{figure} [h]
\begin{center}
\begin{tabular}{c} 
\includegraphics[height=18cm]{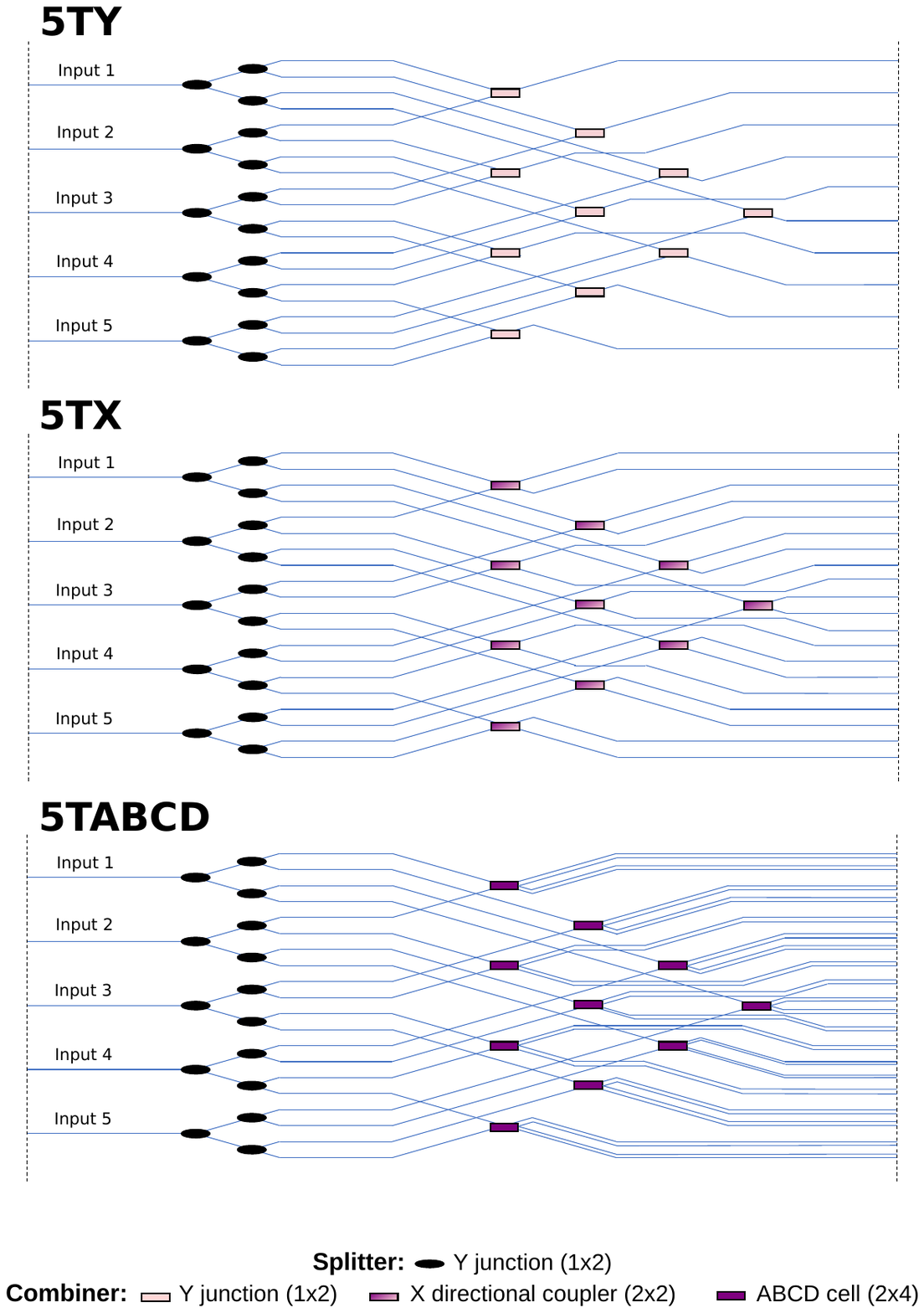}
\end{tabular}
\end{center}
\caption 
{ \label{fig:FIRSTv2BeamCombination} 
Interferometric beam combination scheme for {\color{black}{FIRSTv2 5T-combiner. Three possible PIC designs are presented depending on the type of combiner used: Y junctions (10 outputs, one per baseline), directional couplers (20 outputs) or ABCD {\color{black}{cells}} (40 outputs).}}}
\end{figure}

\begin{figure} [h]
\begin{center}
\begin{tabular}{ll}
\includegraphics[height=7 cm]{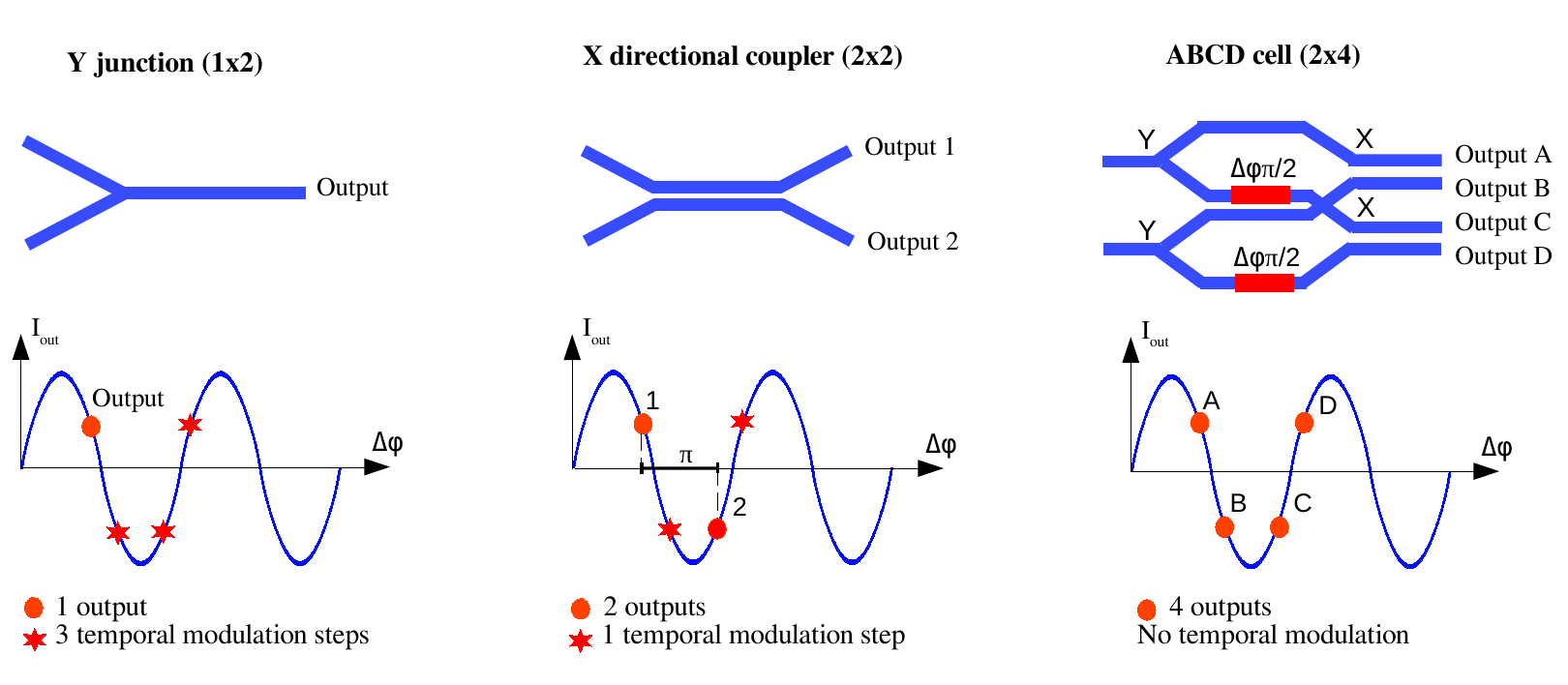}
\end{tabular}
\end{center}
\caption 
{\label{fig:Combiners} {\color{black}{Principle of the three types of combiners considered for FIRSTv2 5T-combiner, and corresponding acquisition process. Y junction and X directional coupler respectively provide one and two measurements of the interference state between the input beams. In both cases, a temporal modulation is required to reconstruct the fringe pattern and assess its phase and amplitude.
An ABCD cell is composed of Y junctions, X directional couplers and achromatic phase shifters. It provides four simultaneous measurements of the interference state at various phase differences, meaning that fringes are reconstructed with a single image.}}}
\end{figure}

{\color{black}{FIRSTv2}} interferometric combination scheme consists in combining the light {\color{black}{of 5 sub-apertures pair by pair}}, as presented in Fig.~\ref{fig:FIRSTv2BeamCombination}. {\elsa{This design is called a 5-telescopes combiner, or 5T-combiner, by keeping the same name convention as PICs developped for long baseline interferometers combining the light of different telescopes.}}
Each of the 5 inputs are split in four thanks to {\color{black}{two cascaded Y junctions}} and are recombined with the other {\elsa{four}} thanks to combiners. {\color{black}{As presented in  Fig.~\ref{fig:FIRSTv2BeamCombination}}}, combiners can be Y junctions (10 outputs {\elsa{in total}}, one per baseline), directional couplers (20 outputs) or ABCD {\color{black}{cells}} (40 outputs). {\color{black}{Fig.~\ref{fig:Combiners} presents an overview of these combiners. In order to properly sample the fringes around the null OPD when using directional couplers and Y junctions, a phase modulation of the input beams is required. This phase modulation is achieved by adding piston to the corresponding segments of the segmented mirror following a 20 steps sequence running at 20\,Hz.}} 
 {\elsa{The ABCD cell provides four measurements of the interference state between the input beams, with four distinct phase differences, allowing the fringe reconstruction with a single image. The use of ABCD cells thus increase the observing time efficiency, however they are more complex to design, as they require {\elsaaa{splitters}}, directional couplers and achromatic phase shifters.}}

\subsection{Specifications for the FIRSTv2 {\elsa{5T-combiner}}}
\label{subsec:spec} 
As {\elsa{explained}} in Sec.~\ref{sec:firstv2}, the FIRSTv2 science case is the detection of accreting protoplanets {\elsa{by differential phase measurement at the H$\alpha$ line}}. It consists in comparing the complex visibility phase at the H$\alpha$  line (at 656.3\,nm, where the protoplanet is brighter) to the complex visibility phase in the continuum (from 630 to 650\,nm and from 660 to 780\,nm, where it is not detected). Therefore, {\elsa{we defined}} PIC specifications {\elsa{depending}} on the wavelength, as shown in Table.~\ref{tab:table3}, {\elsa{with stronger constraints around the H$\alpha$ line}}{\elsaa{, where the signal of interest is expected. }}{\elsa{Indeed, the photon transmission in this spectral channel is critical, while the continuum signal is evaluated on $\sim$\,700 channels when using the whole spectral range.}}{\elsaa{ The specifications concern several aspects:
\begin{itemize}
    \item overall transmission: FIRSTv2 PIC {\color{black}{overall}} transmission must be greater than 50$\%$ {\color{black}{in the $630$ to $780$\,nm spectral band}} and greater than 75$\%$ {\color{black}{around the H$\alpha$ line}}.
    \item insertion or coupling loss: it specifies the amount of energy lost because of the mismatch between the fundamental modes of the PM-630HP fiber, used for the injection, and the PIC waveguide.
    \item Polarization Extinction Ratio (PER): {\color{black}{it characterizes the capability of the PIC building blocks to keep light linearly polarized and propagating along a principal axis. In that case, the PER should be close to 100$\%$. After injecting linearly polarized light along the principal axis, the Polarization Extinction Ratio (PER) corresponds to the ratio between the output power of the light linearly polarized along the principal axis and the output power of the light orthogonally polarized.}}
    \item cross-talk: it defines the amount of {\elsaa{unwanted}} light that gets transferred from one waveguide to another. This specification mainly applies to waveguides crossing each other and does not apply to splitters nor combiners.
\end{itemize}
}}

\begin{table}[h!]
\caption{FIRSTv2 5T PIC {\color{black}{main}} specifications}
\label{tab:table3}
\begin{center}       
\begin{tabular}{|l|l|l|}
\hline
\rule[-1ex]{0pt}{3.5ex} Wavelength & $650$ to $660$\,nm & $6${\color{black}{$3$}}$0$-$650$\,nm and $660$-$780$\,nm \\
\hline
\rule[-1ex]{0pt}{3.5ex} \textbf{{\color{black}{Single-mode}}} & Yes & Yes\\
\rule[-1ex]{0pt}{3.5ex} {\color{black}{Overall}} transmission & $\ge 75\%$ & $\ge 50\%$ \\ 
\rule[-1ex]{0pt}{3.5ex} Insertion loss & $\le 0.13$\,$dB$ & $\le 0.46$\,$dB$ \\
\rule[-1ex]{0pt}{3.5ex} Polarization Extinction Ratio (PER) &  $\ge$ 99$\%$ &  $\ge$ 99$\%$\\
\rule[-1ex]{0pt}{3.5ex} Cross talk & $\le$ 0.1$\%$ & $\le$ 0.1$\%$ \\    
\hline
\end{tabular}
\end{center}
\end{table}

{\color{black}{Specification of FIRSTv2 5T-combiner building blocks, i.e splitters and combiners, are presented in Table.~\ref{tab:tableSPEC}}}. {\elsa{ The internal losses are specified in order to keep the overall transmission above the specified level.}} Transfer rate specifications for splitters and combiners are meant to {\color{black}{evenly distribute the light}} in order to maximize fringe contrast over {\color{black}{the}} spectral band. {\color{black}{Each of the 5 inputs are split in four thanks to two cascaded Y junctions. The Y junction transfer rate should be $50\pm 5\%$ meaning that each of the Y junction outputs contains $50\pm 5\%$ of the input flux. These 5 inputs are recombined with one another thanks to either the Y junction, directional coupler or ABCD cell combiners. For an ABCD cell, the $25\pm5\%$ specification means that each of the four ABCD cell outputs contains close to $25\%$ of the input flux.}}

\begin{table}[h!]
\caption{
{\elsa{Specifications for the FIRST PIC building blocks}}
} 
\label{tab:tableSPEC}
\begin{center}       
\begin{tabular}{|l|l|l|}
\hline
\rule[-1ex]{0pt}{3.5ex} Wavelength & 650 to 660\,nm & $6${\color{black}{$3$}}$0$-$650$\,nm and $660$-$780$\,nm \\
\hline
\rule[-1ex]{0pt}{3.5ex}\textbf{Internal loss:} & & \\
Crossed waveguide & $\le 0.04$\,$dB$ & $\le 0.13$\,$dB$ \\
Y junction & $\le 0.09$\,$dB$  & $\le 0.27$\,$dB$  \\
Directional coupler & $\le 0.22$\,$dB$ & $\le 0.7$\,$dB$\\
\hline
\rule[-1ex]{0pt}{3.5ex}\textbf{Transfer rate:} & & \\
ABCD cell (combiner) & 25 $\pm$ $5\%$ (25:25:25:25) & 25 $\pm$ $15\%$ \\
Y junction (splitter and combiner) & 50 $\pm$ 5$\%$  (50:50) & 50 $\pm$ 15$\%$ \\
Directional coupler (combiner) & 50 $\pm$ 10$\%$ (50:50) & 50 $\pm$ 30$\%$ \\
\hline
\end{tabular}
\end{center}
\end{table}

\subsection{{\color{black}{TEEM Photonics $K_+:Na_+$ ion exchange technology (ioNext)}}}
\label{sec:teems} 
The {\color{black}{FIRSTv2}} PIC is developed {\color{black}{in collaboration with TEEM Photonics\astfootnote{\linkable{https://www.teemphotonics.com/}}, a company}} based in Grenoble, France. TEEM Photonics ioNext technology consists in $K_+:Na_+$ ion exchange in {\elsa{a glass substrate}}. A lithographic mask is {\elsa{used to control the regions where the ion exchange is performed, thus creating gradient-index waveguides featuring a precisely controlled mode field profile and effective index\cite{Tervonen2011}.}} {\color{black}{{\elsaa{At a final stage, a}} glue layer and a counter-blade are deposited on top of the glass substrate. They allow {\elsaa{for}}: 1) a larger front surface {\elsaa{where V-grooves are bonded ; 2) for a protection of }}the waveguide which would {\elsaa{otherwise be directly}} on the surface, and thus sensitive to scratches and dust ; and 3) for the symmetrization of the waveguide mode in order to better {\elsaa{match the modes of}} the inserting and collecting optical fibers.}} 
The SM-630 fiber insertion loss is specified by {\color{black}{TEEM Photonics to be $0.13$\,dB}} over the {\color{black}{single-mode}} spectral band for a $2$\,{\color{black}{$\mu$m}} wide waveguide. {\color{black}Straight waveguide propagation loss is about $0.25$\,{dB/cm}} at $780$\,{\color{black}{nm}}. 

{\elsaa{\subsection{The first 5T-beam combiners prototypes}
\label{subsec:proto}}}
{\elsaaa{Prior to the present work, two 5T PIC prototypes}} {\elsa{fabricated with}} TEEM Photonics ioNext technology were characterized on the FIRSTv2 testbed at the Observatoire de Paris {\cite{barjot2021}}. {\color{black}{These PICs perform the interferometric recombination of 5 sub-apertures beams with two different types of combiners:}} {\elsaaa{one is based on Y junction combiners and is called 5TY, while the other one is based on symmetric directional couplers or X couplers, further detailed in Sec.~\ref{subsec:xcoupler}, and is called 5TX.}}

{\color{black}{As illustrated in Fig.~\ref{fig:FIRSTv2BeamCombination}, Y junction consists in two single-mode waveguides which merge into one output single-mode waveguide. If the fields propagating in input waveguides are in phase (resp. in phase opposition), they are coupled into the fundamental mode (resp. in radiative modes) of the output waveguide, meaning that half of the interferometric signal is lost in radiative modes in a Y junction. {\elsaaa{For that reason, Y junctions were not further investigated and are not part of the present study.}}}}


{\color{black}{The 5T prototypes {\elsaaa{were characterized in terms of throughput and cross-talk}}, as defined in Sec.~\ref{subsec:spec}.{\elsaaa{ Light was injected}} in one of the five input waveguides and the leakage in the other waveguides {\elsaaa{was measured}}. Cross-talk has a mean value of about $1\%$ in both PIC prototype but can reach, for some inputs, $10\%$ for the 5TX PIC and $20\%$ for the 5TY PIC. The throughput was estimated to about 15$\%$ (5TY) to 30$\%$ (5TX)\cite{barjot2021}. This low throughput is mainly due to non-optimized combiners and bend curvature radii. {\elsaaa{Further developments are thus needed to reach the specifications presented in Sec.~\ref{subsec:spec}, motivating the work presented in the following sections.}}}} 
{\color{black}{The methodology adopted to develop the PICs is {\elsaaa{indeed}} an iterative process. A numerical model of the waveguides manufactured with the ioNext technology is {\elsaaa{required to feed the}} BeamPROP modeling software. Combiners are designed and optimized based on the numerical simulations of their performance. {\elsaaa{Once manufactured,}} these combiners are characterized in the laboratory to refine the numerical model {\elsaaa{and tune the design parameters for the next manufacturing run.}}}}

{\elsaa{\section{Fundamental parameter estimation for waveguide modeling\label{sec:3model}}}}

{\color{black}{In order to model the waveguides built with the ioNext technology and design FIRST PIC's combiners, fundamental parameters estimation of standard, i.e straight, waveguides is needed. In particular, the fabrication process, i.e K+ NA+ ion exchange in glass, creates a diffused index profile in a glass substrate which needs to be recovered. This section presents the laboratory measurements of single-mode spectral range and mode field diameter of ioNext standard straight waveguides which are used to defined the 3D diffused index profile of the waveguide on the BeamPROP software.}}

\subsection{Single-mode spectral range laboratory measurement}
\label{subsec:lab_smode}
\begin{figure} [h]
\begin{center}
\begin{tabular}{c} 
\includegraphics[height=8.5cm]{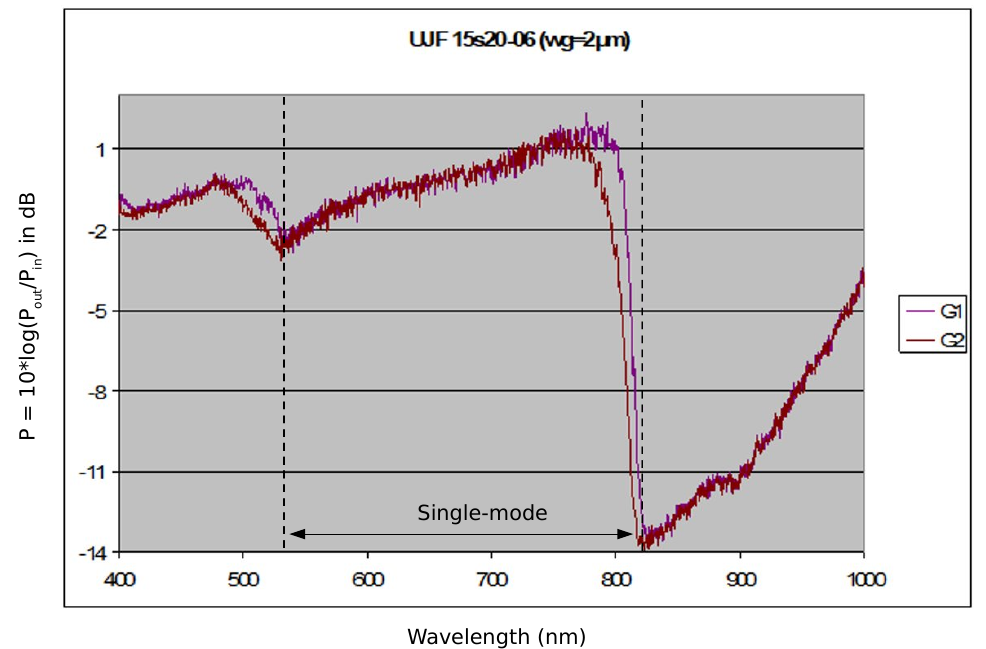}
\end{tabular}
\end{center}
\caption 
{ \label{fig:SingleModeRange} {\color{black}{Single-mode spectral band measurement}} for {\color{black}{two straight waveguide {\elsa{samples}} $G1$ and $G2$}} built with a 2\,{\color{black}{$\mu$m}} wide lithographic mask. {\color{black}{The throughput $P=10\cdot\log(P_{out}/P_{in})$ is estimated with $P_{in}$ the fiber-to-fiber reference power measurement. $P_{out}$ is measured by inserting the straight waveguide of the photonic integrated circuit between these two single-mode fibers.}}}
\end{figure}

 {\color{black}{Figure~\ref{fig:SingleModeRange} presents single-mode spectral range measurement performed by TEEM Photonics engineers.}}{\color{black}{ The throughput $P=10\cdot\log(P_{out}/P_{in})$ is measured for two straight waveguides {\elsa{test samples}} $G1$ and $G2$. The white light source {\elsa{is fibered and the measurement is performed with a fibered spectrometer. The}} input power $P_{in}$ is thus estimated by connecting {\elsa{directly the source fiber to the spectrometer fiber}}. The output power $P_{out}$ is measured by {\elsa{inserting}} the photonic integrated circuit straight waveguide between these two fibers. 

For wavelengths greater than $820$\,nm, the input light is not guided by the waveguide. When the wavelength decreases and as soon as the input light gets guided through the fundamental mode (m=0) of the waveguide, $P_{out}$ increases explaining the peak at the cutoff wavelength around $820$\,nm. In the single-mode range, the coupling {\elsa{efficiency}} between the fiber and the waveguide fundamental modes decreases with the wavelength. 
An increasing amount of the input light is coupled into the waveguide second mode (m=1) and immediately lost because of the low confinement of this second mode. As soon as the wavelength gets small enough, the energy coupled in the second mode becomes confined and is transmitted through the waveguide: a second peak in the output power appears at this second cutoff wavelength of about $530$\,nm.}} The {\color{black}{measured single-mode spectral range {\elsa{extends between these two cut-off wavelengths,}} approximately between $530$ to $820$\,nm}} for $2$\,$\mu$m wide waveguides.

{\elsa{It can be noted that}} the throughput value should be lower than 0\,dB because this passive straight waveguide should only insert losses. This could be explained by the measurement noise or by {\elsa{losses induced by the }}fiber-to-fiber alignment, {\elsa{performed inside a standard connector, while}} the fiber-to-waveguide alignment {\elsa{is precisely tweaked, potentially}} leading to {\elsa{an underestimation of the}} input power measurement. Also, no information is given {\elsa{regarding}} the input polarization and as the waveguides are known to be birefringent, a different single-mode spectral range is to be expected between {\color{black}{p- and s-polarized light}}.

\subsection{Mode field diameter laboratory measurement}
\label{subsec:lab_mfd}
\begin{figure} [h!]
\begin{center}
\begin{tabular}{c} 
\includegraphics[height=12 cm]{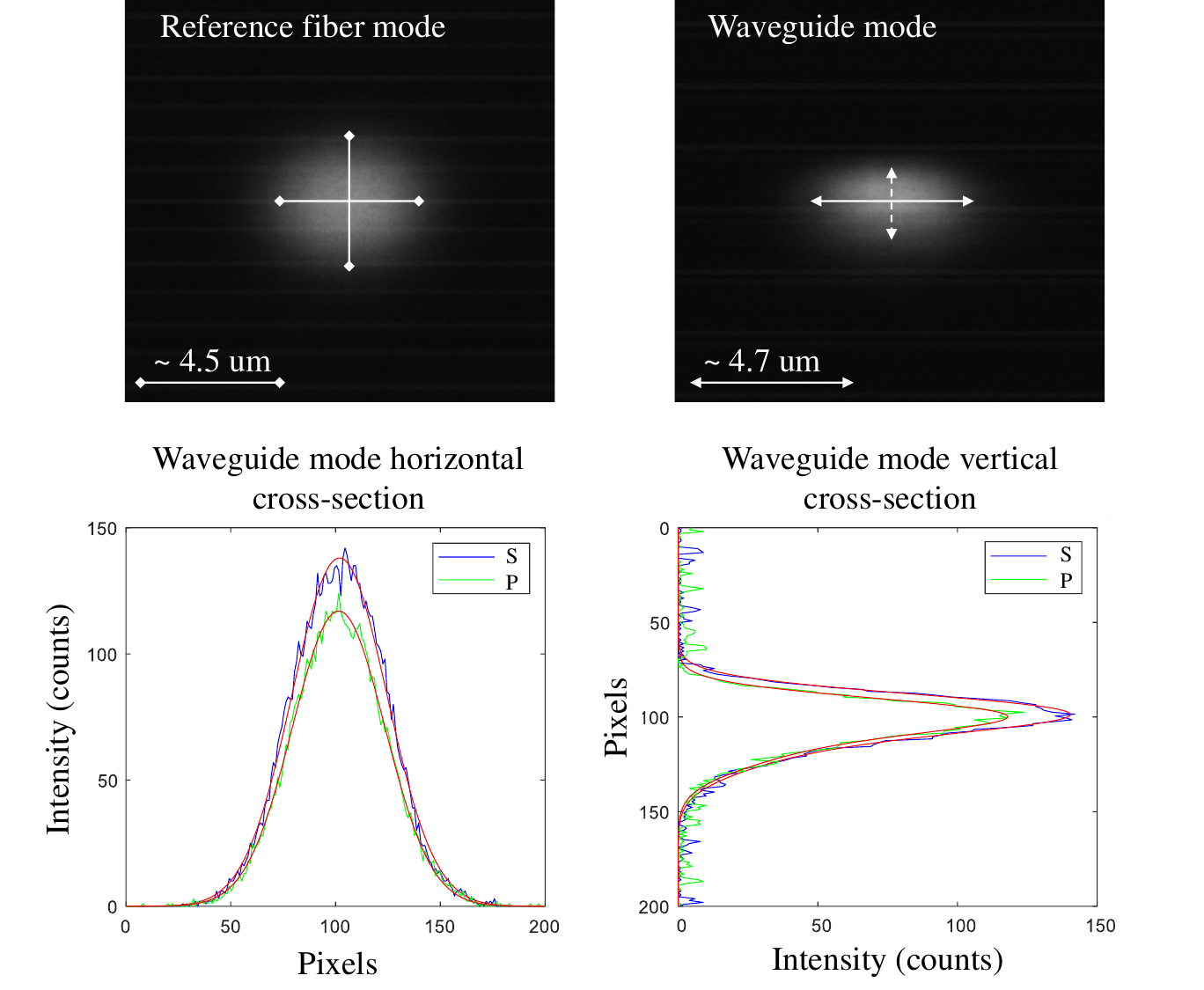}
\end{tabular}
\end{center}
\caption 
{ \label{fig:ModeFieldDiameters} TEEM Photonics waveguide mode field diameters (MFD) at $635$\,nm. 
\textbf{Top left:} PM-630-HP reference fiber mode intensity recorded for a {\color{black}{p-polarized}} light injection. \textbf{Top right:} TEEM Photonics waveguide fundamental mode intensity recorded for a {\color{black}{p-polarized}} light injection. \textbf{Bottom left:} intensity profile for a horizontal cross-section of the top right TEEM Photonics waveguide mode field diameter image (along the plain arrow). \textbf{Bottom right:} intensity profile for a vertical cross-section of the top right TEEM Photonics waveguide mode field diameter image (along the dashed arrow). {\elsa{The red curves show the Gaussian best fit model, leading to the estimation of the MFD.}}}
\end{figure}

{\color{black}{TEEM Photonics waveguide mode field diameter (MFD) measurements were performed for {\color{black}{p- and s-polarized}} light injection}}. {\color{black}{In this particular experiment, p-polarized (resp. s-polarized) electric field lies in a plane parallel (resp. orthogonal) to the plane of the PIC.}} {\color{black}{The Thorlabs PM-630-HP reference fiber and TEEM Photonics waveguide outputs are imaged with a x40 objective on a Thorlabs CDD.}} {\elsa{The MFD are calibrated for {\color{black}{p- and s-polarized light}} injection thanks to a reference polarization maintaining fiber with a known $4.5$\,$\pm$\,$0.5$\,$\mu$m MFD.}} The {\color{black}{MFD measurements}} for a horizontal and vertical {\color{black}{cross-section}} of TEEM Photonics waveguide mode are presented in Fig.~\ref{fig:ModeFieldDiameters} and Table.~\ref{tab:table5}.

\begin{table}[h]
\label{tab:table5}
\begin{center}       
\begin{tabular}{|l|l|l|l|}
\hline
\rule[-1ex]{0pt}{3.5ex} Polarization & Mode width & Mode upper height & Mode lower height \\
\hline
\rule[-1ex]{0pt}{3.5ex} {\color{black}{s-polarized}} & 4.81\,$\pm$\,0.64\,$\mu$m & 0.74\,$\pm$\,0.17\,$\mu$m & 1.12\,$\pm$\,0.42\,$\mu$m \\
\hline
\rule[-1ex]{0pt}{3.5ex} {\color{black}{p-polarized}} & 4.68\,$\pm$\,0.61\,$\mu$m & 0.65\,$\pm$\,0.15\,$\mu$m & 1.07\,$\pm$\,0.29\,$\mu$m\\
\hline
\rule[-1ex]{0pt}{3.5ex} Mean & 4.75\,$\pm$\,0.69\,$\mu$m & 0.70\,$\pm$\,0.21\,$\mu$m & 1.10\,$\pm$ 0.38\,$\mu$m\\
\hline
\end{tabular}
\end{center}
\caption{\label{tab:table5}Polarized mode field diameters measured at $635$\,{\color{black}{nm}}. Mode width corresponds to the $1/e^2$ width of the mode horizontal {\color{black}{cross-section presented on the bottom left graph Fig.~\ref{fig:ModeFieldDiameters}}}. Because the mode is asymmetric in the vertical direction, two measurements are performed in this direction: the mode upper (resp. lower) height corresponds to the upper (resp. lower) $1/e^2$ half-height of the mode presented on the bottom right graph Fig.~\ref{fig:ModeFieldDiameters}.} 
\end{table}

{\color{black}{MFD measurements reveal the mode asymmetry induced by the fabrication process: the ion exchange is performed at the glass substrate surface {\elsa{meaning that the}} waveguides are not buried into the glass, inducing a form birefringence in spite of the glue layer and the counter-blade being used for symmetrization. Currently}}, the precision of the mode field diameter measurement is not sufficient to define a polarization-dependent model of the ioNext technology {\elsa{that could be taken into account in the BeamPROP simulation to refine the design}}. {\elsa{As a consequence}}, a non-polarized mode field diameter is defined by taking the mean values of polarized mode field dimensions. Effective indexes and throughput in both polarizations will be further investigated thanks to the characterization of new PICs, see Sec.~\ref{subsec:wafer}.
\subsection{Modeling of TEEM Photonics waveguide diffused index profile {\elsa{with}} the BeamPROP software}
\label{subsec:model_index}

\begin{figure} [h!]
\begin{center}
\begin{tabular}{c} 
\includegraphics[height=5.5 cm]{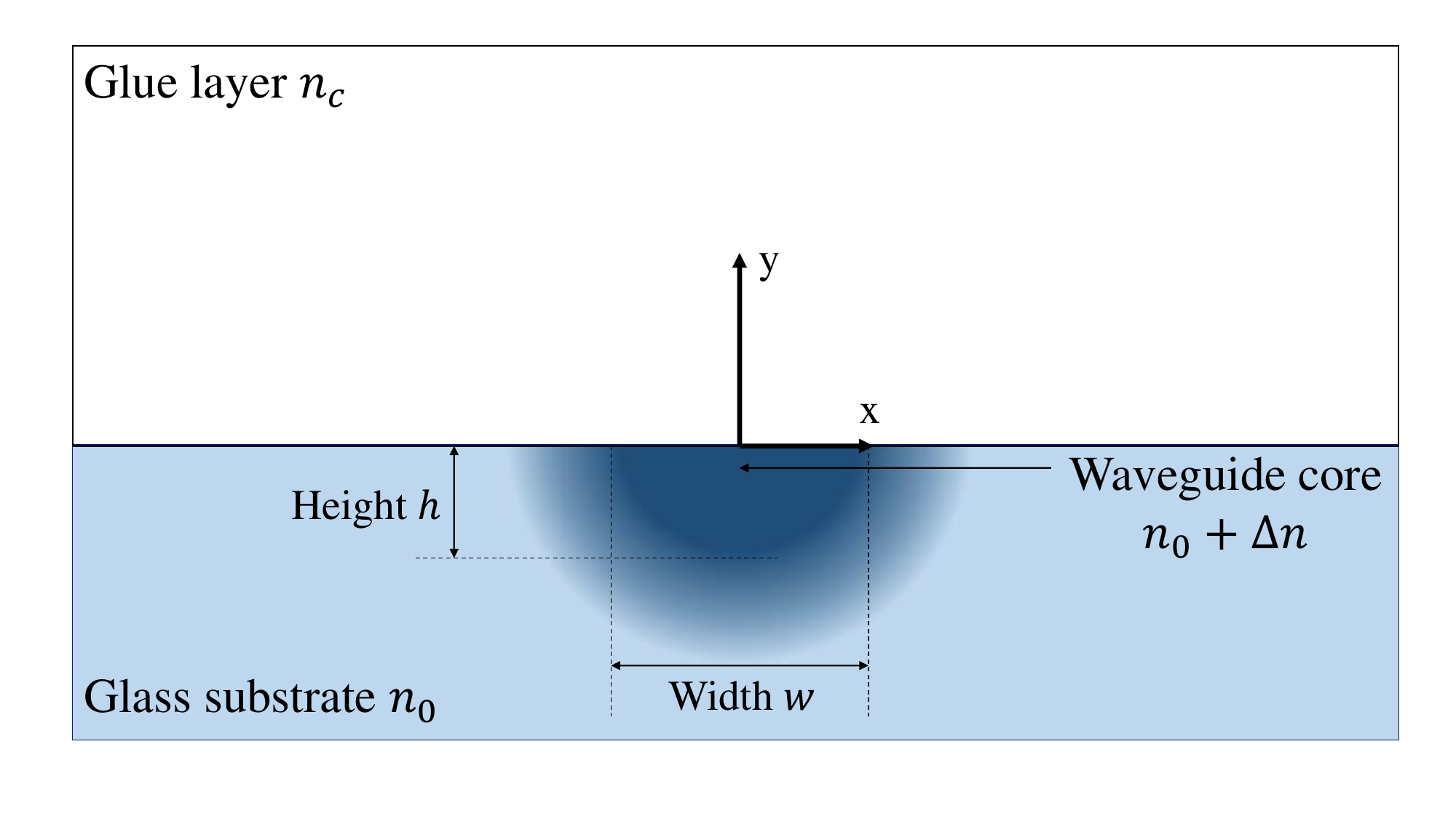}
\end{tabular}
\end{center}
\caption 
{ \label{fig:BeamPROPModel} {\color{black}{Parametrization of the 3D diffused index profile cross-section}} \textit{(adapted from BeamPROP software user guide)}}
\end{figure}

Based on the {\color{black}{single-mode spectral range}} and non-polarized mode field diameter measurements, a 3D diffused index profile {\color{black}{of TEEM Photonics waveguide}} is derived on the BeamPROP {\color{black}{modeling}} software\astfootnote{\linkable{https://www.synopsys.com/photonic-solutions/rsoft-photonic-device-tools/passive-device-beamprop.html}}. {\color{black}{The 3D diffused index profile parameters are}} presented in Fig.~\ref{fig:BeamPROPModel}. 

The glue layer is about $20$\,$\mu$m {\color{black}{thick in the $y>0$ region and is considered as semi-infinite. Its refractive index is $n_c=1.49$ at $635$\,nm. The width of the lithographic mask used for the diffusion process is $w$.}} The diffusion length in the horizontal (resp. vertical) direction is $h_x$ (resp. $h_y$). {\color{black}{The maximum refractive index difference between the glass substrate and the waveguide core produced by the diffusion process is $\Delta n$.}}
The waveguide diffused index profile cross-section $n(x,y)$ {\color{black}{in the glass substrate ($y<0$)}} is defined by: 
\begin{equation}
\label{eq:equation1}
n(x,y<0)=n_o+[\Delta n\cdot g(x)\cdot f(y)]^\gamma,
\end{equation}
{\elsa{with $n_0=1.52$ at $635$\,nm the {\color{black}{glass substrate $(y<0)$}} refractive index, and where the functions $g(x)$ and $f(x)$ describe the region where the diffusion takes place:}} 
\begin{equation}
\label{eq:equation2}
g(x)=\frac{1}{2}\left\{erf[(\frac{w}{2}-x)/h_x)]+erf[(\frac{w}{2}+x)/h_x]  \right\},
\end{equation}
\begin{equation}
\label{eq:equation3}
f(y)=exp(\frac{-y^2}{h_y^2}).
\end{equation}
For the sake of simplicity {\color{black}{in this preliminary model}}: 1) the diffusion process is assumed to be equivalent in both x and y directions, i.e. $h_x=h_y=h$ ; 2) the relation between the ion concentration and the index $n(x,y)$ is supposed to be linear, i.e $\gamma=1$ in Eq.~\ref{eq:equation1}. {\color{black}{{\elsa{In practice, the}} diffusion length $h$ and maximum refractive index difference $\Delta n$ parameters are optimized thanks to laboratory measurements as reported in Sec.~\ref{subsec:lab_smode} and Sec.~\ref{subsec:lab_mfd}.}}

{\elsaa{\section{Design and optimization of three types of combiners\label{sec:design}}}}

{\elsaa{Different types of combiners are considered in this section: symmetric and asymmetric directional couplers, and ABCD cells (composed of Y junction, directional couplers and achromatic phase shifters)}}. {\color{black}{Laboratory characterization of symmetric directional couplers and design of asymmetric directional couplers and ABCD cells are presented.}} {\elsaa{Directional couplers, also called X directional couplers according to their shape, recombine the light from two input waveguides and lead to two outputs, corresponding to the coupling of the two waveguides with a phase difference of 0 and $\pi$. ABCD cells involve a more complex design, including splitters, phase shifters and directional couplers. They provide four outputs which correspond to the coupling of the two input waveguides with a phase difference of 0, $\pi/2$, $\pi$ and $3\pi/2$, see Fig.\ref{fig:Combiners}.}}

{\elsaa{\subsection{Symmetric directional couplers}}}
\label{subsec:xcoupler}
\subsubsection{Theoretical model}
\label{subsec:xcoupler_theo}
\begin{figure} [h]
\begin{center}
\begin{tabular}{l}
\includegraphics[height=6cm]{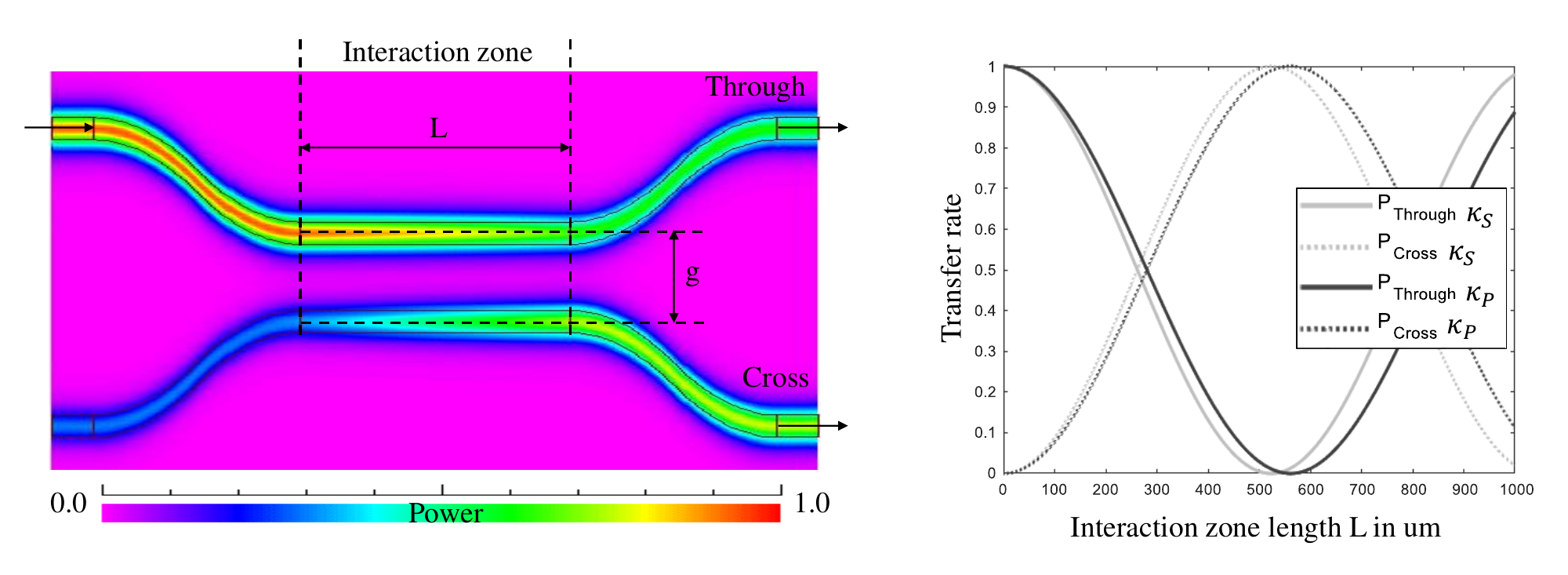}
\end{tabular}
\end{center}
\caption 
{ \label{fig:DirectionalCouplerTheory} \textbf{Left:} Standard geometry of a symmetric directional coupler, {\elsa{with the interaction zone defined by its length $L$ and gap distance $g$ between the coupling waveguides}}. \textbf{Right:} Theoretical {\color{black}{normalized output power}} as a function of the interaction zone length $L$ for a symmetric directional coupler. The {\color{black}{mode}} coupling coefficient $\kappa$ depends on wavelength and on polarization {\color{black}{(i.e $\kappa_S$ $\neq$ $\kappa_P$)}} for birefringent waveguides.}
\end{figure}

The directional coupler standard geometry is presented in Fig.~\ref{fig:DirectionalCouplerTheory}{\elsa{: two waveguides come very close to one another, inducing a coupling between the two by evanescent waves in the so-called interaction zone. For simulation and characterization purposes,}} the waveguide called "Through" (resp. "Cross") is the waveguide in which the light is injected (resp. not injected). The interaction zone is defined by {\color{black}{its length $L$ and the gap distance $g$ between the waveguides}}. {\color{black}{The directional coupler is called symmetric if both waveguides in the interaction zone have the same geometrical properties.}} 

The {\color{black}{normalized}} output powers $P_{\rm{through}}$ and $P_{\rm{cross}}$ of a {\color{black}{lossless}} directional coupler\cite{labeyephd} are given by: 
\begin{equation}
\label{eq:equation4}
P_{\rm{through}} = 1-P_{\rm{cross}} = 1-(\kappa^2/\Delta^2)\cdot sin^2(\Delta \cdot L) 
\end{equation}

with $\kappa$ the mode coupling coefficient and $\Delta$ defined as:  

\begin{equation}
\label{eq:equation5}
\Delta=\sqrt{(\frac{\beta_{\rm{through}}-\beta_{\rm{cross}}}{2})^2+\kappa^2}
\end{equation}
with $\beta_{\rm{through}}$ and $\beta_{\rm{cross}}$ the propagation constants of the {\color{black}{directional coupler}} waveguides. {\color{black}{In Equation\,\ref{eq:equation4}, $F=\kappa^2/\Delta^2$ corresponds to the maximum power that can be transferred from one waveguide to the other. For a symmetric directional coupler, as both waveguides in the interaction zone have the same geometrical properties, propagation constants are equal ($\beta_{\rm{through}}\,=\,\beta_{\rm{cross}}$). Therefore F is equal to 1, meaning that all the power in one waveguide can be transferred into the other one, if the interaction zone length is properly chosen. The shortest length of the interaction zone for which the power is entirely transferred from one waveguide to the other is called $L_{0:100}$. Equation\,\ref{eq:equation4} shows that the transferred power $P_{\rm{cross}}$ is periodic with a period proportional to $2L_{0:100}$, with $L_{0:100}=\frac{\pi}{2\Delta}$.}}
The directional coupler transfer rate {\color{black}{$A:B$ (with $A+B=1$)}} is defined as the proportion of the output flux in each output : 

\begin{equation}
\label{eq:equation6}
\frac{P_{\rm{through}}}{P_{\rm{through}}+P_{\rm{cross}}} \textrm{ : } \frac{P_{\rm{cross}}}{P_{\rm{through}}+P_{\rm{cross}}}
\end{equation}

The specification for the PIC {\color{black}{directional coupler}} is to {\elsa{transmit}} {\color{black}{50 $\pm$ 10$\%$ of the light in each output over the $600$ to $800$\,nm spectral band, i.e to have an achromatic}} transfer rate of 50:50. {\color{black}{In order to get a 50:50 transfer rate, the shortest interaction zone length $L_{50:50}$ is $\frac{L_{0:100}}{2}$. Other solutions $L_{50:50,k}$ for the interaction length are derived from Eq.~\ref{eq:equation4}:}}

\begin{equation}
\label{eq:equation7}
{\color{black}{L_{50:50,k}=\frac{1}{\Delta}(\frac{\pi}{4}+k\frac{\pi}{2}) = L_{0:100}(\frac{1}{2}+k) \quad \textrm{with} \quad k\in \mathbb{N}}}
\end{equation}
As the interaction zone length $L_{0:100}$ depends on the wavelength, the transfer rate of a symmetric coupler is {\elsa{necessarily}} chromatic\cite{labeyephd}. {\elsa{However, a directional coupler can be designed with asymmetric waveguides, leading to }}$\beta_{\rm{through}}\neq\beta_{\rm{cross}}$. {\elsa{The asymmetry can be tuned in order to compensate for the transfer rate chromaticity, as explained in Sect.\,\ref{subsec:asymx}.}} 
\subsubsection{Laboratory characterization}
\label{subsec:xcoupler_lab}
Three symmetric directional couplers {\color{black}{were manufactured by TEEM Photonics using the ioNext technology presented in Sec.~\ref{sec:teems}}}. 
{\color{black}{The symmetric directional couplers all have a gap $g = 2$\,$\mu$m and three different interaction zone lengths: $L=50$\,$\mu$m, $100$\,$\mu$m or $150$\,$\mu$m. Measurements are performed using a P2-830A fiber connected to a white SLED source to inject non-polarized light into the PIC{\color{black}{'s}} symmetric directional couplers. A x4 objective is used so that the light from the PIC output feeds an OceanOptics spectrometer. A linear polarizer, located between the objective and the spectrometer, selects the {\color{black}{p- or s-polarized light. In this particular case, p-polarized (resp. s-polarized) electric field lies in a plane parallel (resp. orthogonal) to the plane of the PIC.}}}}

\begin{figure} [h!]
\begin{center}
\begin{tabular}{ll}
\includegraphics[height=14 cm]{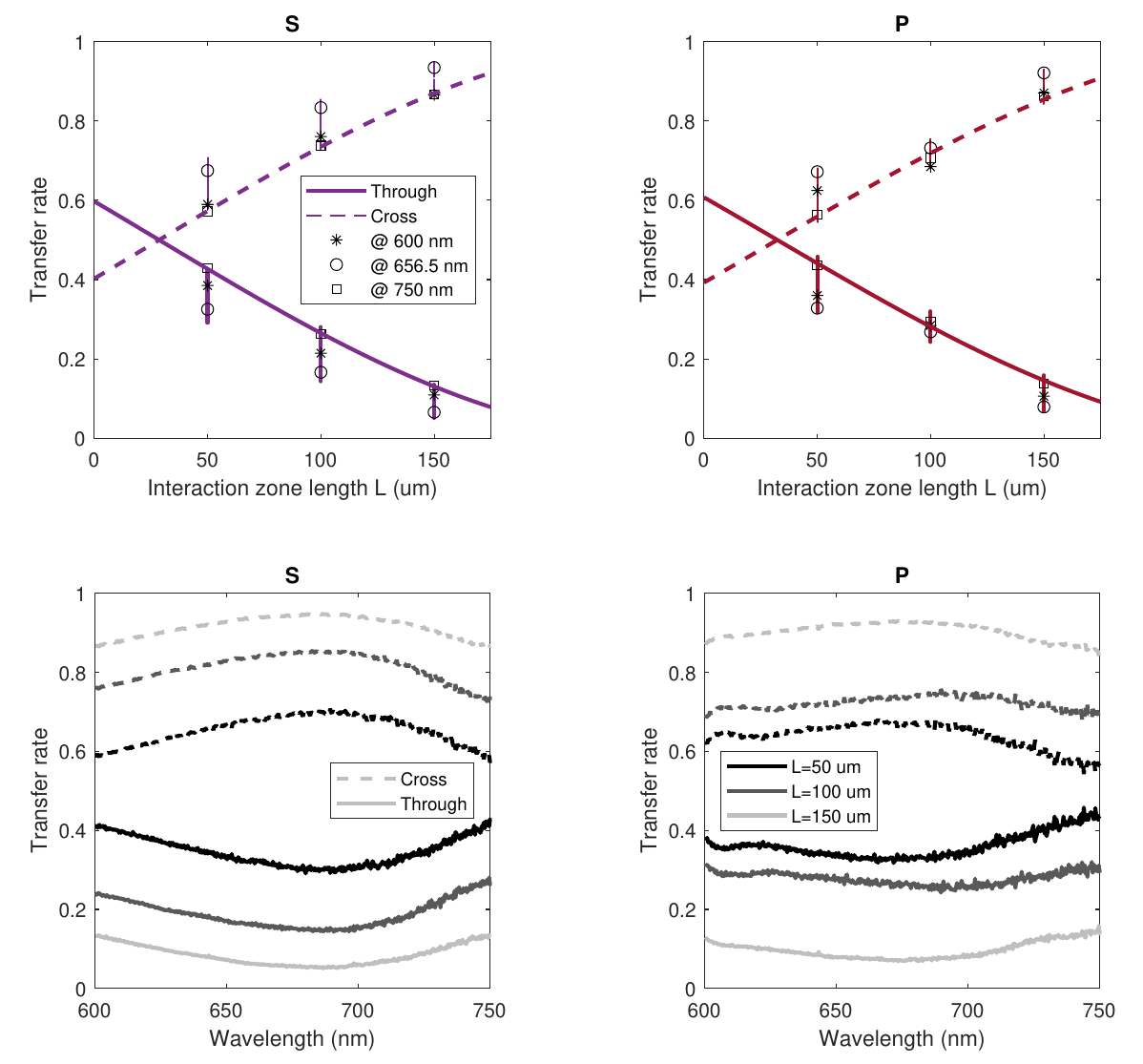}
\end{tabular}
\end{center}
\caption 
{\label{SymmetricDirectionalCoupler} {\elsa{Symmetric directional couplers experimental characterization results.}} \textbf{Top:} Transfer rate for {\color{black}{p- and s-polarized light}} as a function of the interaction zone length $L${\elsa{, experimentally evaluated for three symmetric directional couplers of lengths: $L=50$\,$\mu$m, $100$\,$\mu$m and $150$\,$\mu$m.}} Vertical lines show the maximum variation of the transfer rate over the {\color{black}{$600$ to $750$\,nm}} spectral band while some particular wavelengths are highlighted with markers. The curves correspond to {\elsa{the best fit model as presented in Sec.~\ref{subsec:xcoupler_theo}. \textbf{Bottom:} Transfer rate for {\color{black}{p- and s-polarized light}} as a function of wavelength for the three considered interaction zone lengths.}}}
\end{figure}

{\elsa{Transfer rate measurements, as defined by Eq.~\ref{eq:equation6}, are presented}} in Fig.~\ref{SymmetricDirectionalCoupler}, for both {\color{black}{p- and s-polarized light}}. The {\elsa{objective is twofold: 1) find the interaction zone length $L_{50:50}$ to get a 50:50 transfer rate and 2) investigate the wavelength dependence of the transfer rate}} over {\color{black}{the $600$ to $750$\,nm}} spectral band.

{\elsa{The results}} {\color{black}{presented in}} Fig.~\ref{SymmetricDirectionalCoupler} confirm that the coupling coefficient depends on polarization, i.e waveguides are birefringent. {\elsa{Birefringence}} is not an issue for our application as long as there is no cross-talk between {\color{black}{{\color{black}{p- and s-polarized}} light (i.e no polarization cross-talk)}} {\elsa{that would reduce the fringe contrast}}. {\color{black}{In the {\elsa{present experiment, the input light is not polarized, such that polarization cross-talk cannot be assessed, but it will be the focus of a future study}}.}} {\elsa{Despite the polarization dependency, our experimental measurements show}} that a compromise can be found for the interaction zone length $L_{50:50}$ which is situated between 20 and 40\,$\mu$m {\elsa{for both}} polarizations. 
{\elsa{Furthermore, it is noticeable that}} these symmetric directional couplers are not highly chromatic, in particular for the P polarization. {\elsaa{Indeed, the relative standard deviations are smaller than 5.5$\%$ over the $600$-$700$\,nm range}}, as shown in Table.~\ref{tab:tableChrom}.

\begin{table}[h!]
\label{tab:tableChrom}
\begin{center}       
\begin{tabular}{|c|l|l|l|}
\hline
\rule[-1ex]{0pt}{3.5ex} Interaction zone length $L$ & $50$\,$\mu$m & $100$\,$\mu$m & $150$\,$\mu$m \\
\hline
\rule[-1ex]{0pt}{3.5ex} RSD for {\color{black}{s-polarized light}} & 5.5$\%$ & 4.3$\%$ & 2.8$\%$ \\
\hline
\rule[-1ex]{0pt}{3.5ex} RSD for {\color{black}{p-polarized light}} & 5.0$\%$ & 2.2$\%$ & 2.4$\%$ \\
\hline
\end{tabular}
\end{center}
\caption{\label{tab:tableChrom} \color{black}{Relative standard deviation (RSD) of spectral transfer rates shown in Fig.~\ref{SymmetricDirectionalCoupler} {\elsaa{bottom}}. The relative standard deviation is defined as the ratio between the standard deviation and the mean.}}
\end{table}


\pagebreak

\subsection{Asymmetric directional couplers}
\label{subsec:asymx}

\begin{figure} [h]
\begin{center}
\begin{tabular}{ll}
\includegraphics[width=\textwidth]{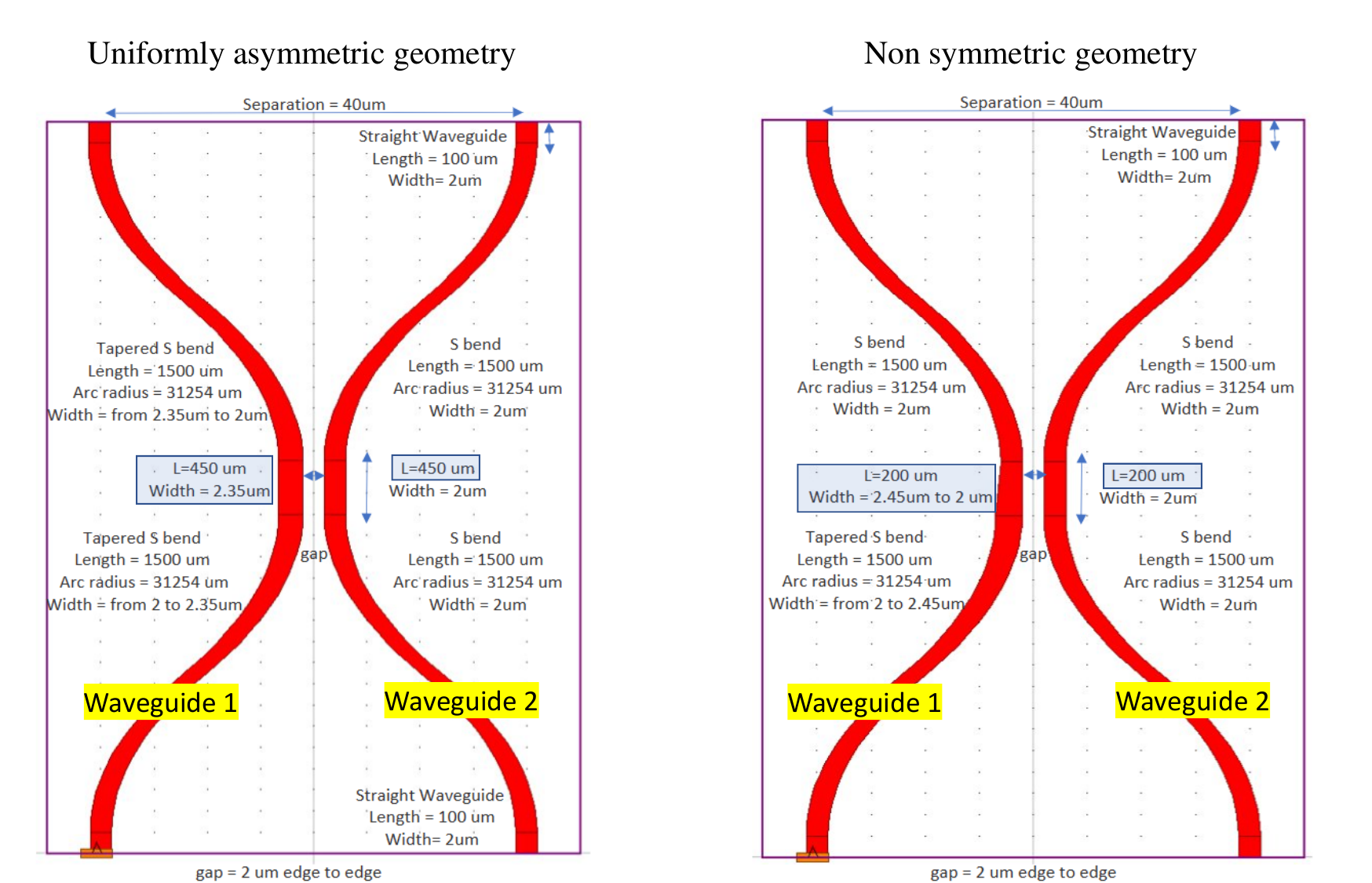} 
\end{tabular}
\end{center}
\caption 
{\label{fig:AsymDirCouplerGeometry} {\color{black}{Two}} asymmetric directional coupler geometries optimized {\color{black}{to meet the transfer rate specification of 50:50 with an accuracy of $\pm 10$ over the 600 to 750\,nm spectral band.}} 
}
\end{figure}

{\elsa{\subsubsection{Asymmetric geometries}}}
\label{subsec:asym_geo}
This section presents two asymmetric directional couplers geometries defined as uniformly asymmetric (UA) and non-symmetric (NS)\cite{Takagi1992} as shown in Fig.~\ref{fig:AsymDirCouplerGeometry}.

A uniformly asymmetric directional coupler is composed of one waveguide wider than the other by an amount $dw$ in the interaction zone. For non-symmetric directional coupler, one waveguide is wider than the other one in the interaction zone {\elsa{and}} it is {\elsa{additionally}} tapered {\elsa{along the interaction zone. This means that the width is continuously varying from one starting value $w$ to a final value $w+dw$.}} 

\subsubsection{Optimization based on BeamPROP simulations}
\label{subsec:asym_opti}
\begin{figure} [h]
\begin{center}
\begin{tabular}{ll}
\includegraphics[height=6cm]{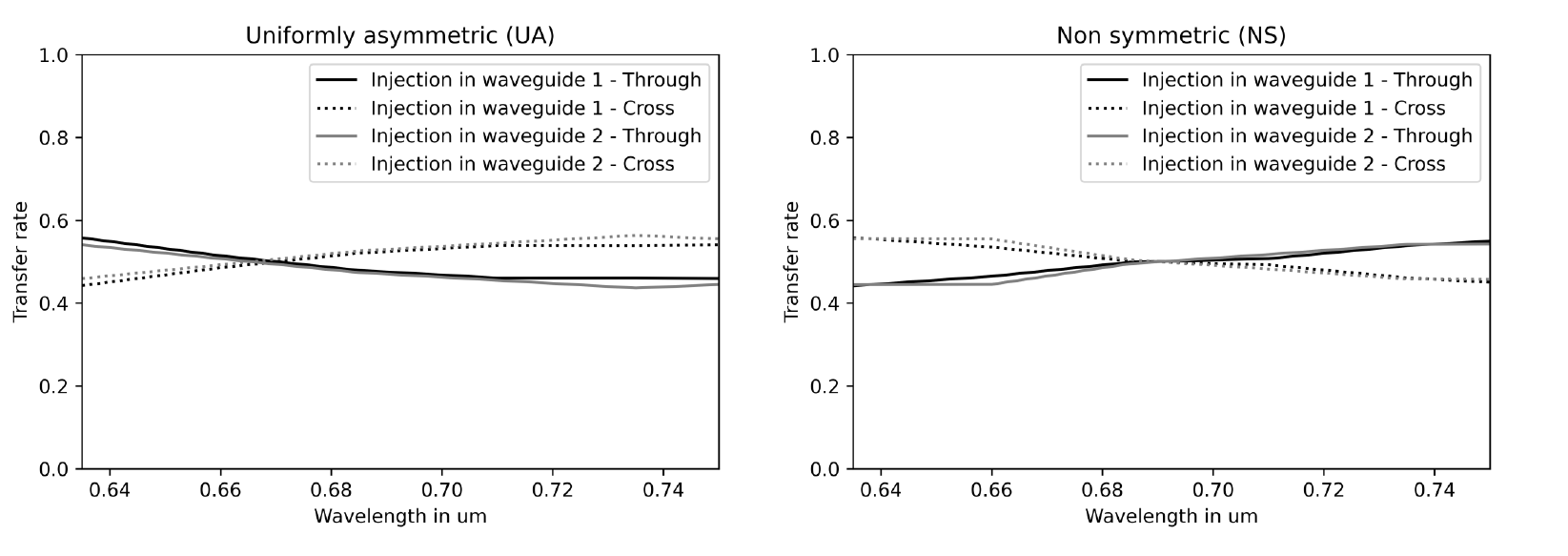}
\end{tabular}
\end{center}
\caption 
{\label{fig:SimulatedTransferRate} {\elsa{Simulated transfer rates (defined by Eq.~\ref{eq:equation6}) as a function of wavelength}}, for injection in each waveguide of the coupler, i.e waveguide 1 or 2 {\elsa{as labelled}} in Fig.~\ref{fig:AsymDirCouplerGeometry}. Normalized output fluxes are plotted for both output arms {\elsa{("Through" or "Cross"): }}for injection in waveguide 1 (resp. waveguide 2), the "Through" waveguide is waveguide 1 (resp. waveguide 2). 
}
\end{figure}

{\elsa{The parameter space ($dw$,$L$) is probed to define the optimal set of parameters for each geometry.}} {\color{black}{The optimal solutions found for uniformly asymmetric and non-symmetric directional couplers are detailed in Fig.~\ref{fig:AsymDirCouplerGeometry}. For both types of asymmetric directional couplers, the spectral}} transfer rates are {\elsa{computed and plotted in Fig.~\ref{fig:SimulatedTransferRate}. For each coupler geometry, the transfer rate is evaluated when injecting in waveguide 1 (“Through” is “waveguide 1” in this case) or waveguide 2 (“Through” is “waveguide 2” in this case). Our simulation results show that the behavior of the coupler weakly depends on the input waveguide.}} 

\begin{figure} [h]
\begin{center}
\begin{tabular}{ll}
\includegraphics[height=10 cm]{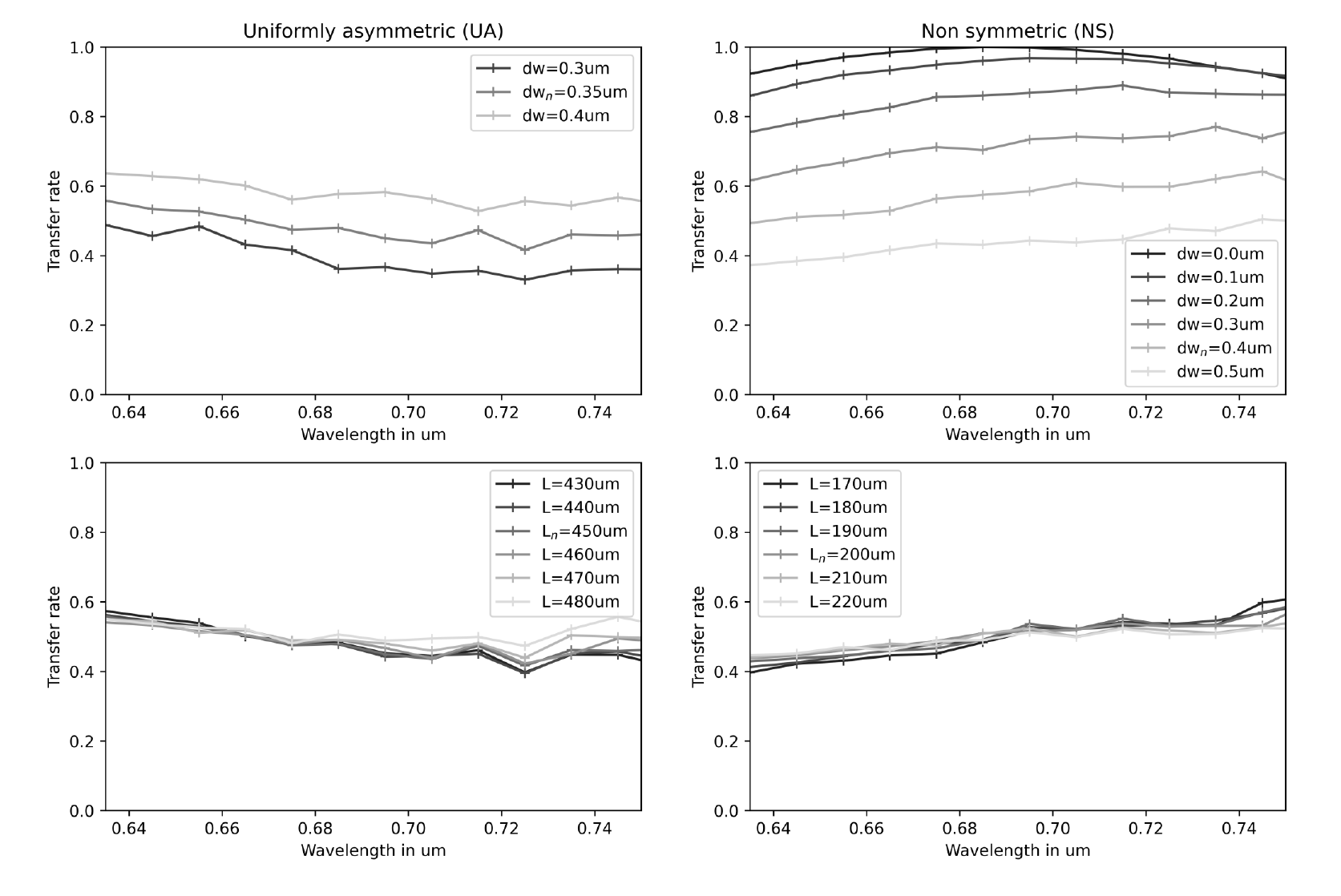}
\end{tabular}
\end{center}
\caption 
{\label{fig:SimulatedTransferRateBis} {\color{black}{Simulated transfer rates for UA {\elsa{(left)}} and NS {\elsa{(right)}} directional couplers, and for various parameter doublets ($dw$,$L$) around the {\elsaa{nominal}} parameter doublets ($dw_n$,$L_n$). For the sake of clarity, transfer rates are shown for injection in waveguide 1 (“Through” = “waveguide 1”) and only the fraction of the output flux in the "Through" waveguide is plotted (i.e $P_{\rm{through}}/({P_{\rm{through}}+P_{\rm{cross}}})$). 
 \textbf{Top:} Transfer rate as a function of wavelength for different waveguide differential widths in the interaction zone.
\textbf{Bottom:} Transfer rate as a function of wavelength for different interaction zone lengths $L$.
}}}
\end{figure}

{\elsa{A study has also been carried out for tolerancing purposes, by computing the spectral transfer rate for various}} differential widths and interaction zone lengths around the {\elsaa{nominal}} parameters doublet ($dw_n$,$L_n$).
{\elsa{The simulation results are presented in Fig.~\ref{fig:SimulatedTransferRateBis}, showing that the differential width accuracy is critical, while a change of a few tenths of microns of the interaction zone length has no significant impact.}} 


{\elsaa{\subsection{ABCD cell combiners}}}
\label{sec:abcd}

{\elsaa{Y junctions and directional couplers are the simplest combiners that can be designed. However, they provide only one or two measurement points per interfering pairs, which is not sufficient to determine the fringe phase and amplitude. In this case, an additional temporal phase modulation is required, to reach a minimum of four fringe samples. In the FIRST instrument, a segmented mirror located before the PIC is used to vary the relative piston between the beams, pair by pair. For a large number of baselines, the data acquisition procedure thus becomes long and complex, and sensitive to phase perturbations occurring from one frame to the other.}}
A way to increase the data  acquisition efficiency is to use another type of combiner called an ABCD cell as presented in the following section.

\subsubsection{Theoretical model}
\label{subsec:abcd_theo}

\begin{figure} [h]
\begin{center}
\begin{tabular}{l}
\includegraphics[height=2.75 cm]{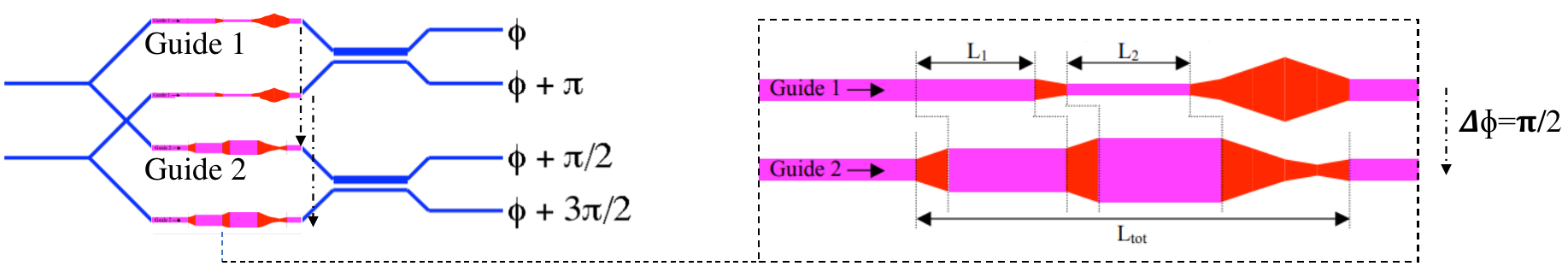}
\end{tabular}
\end{center}
\caption 
{\label{fig:ABCDTheory} ABCD beam {\color{black}{combination}} scheme adapted from P. Labeye's PhD thesis and Benisty et al. 2009\cite{labeyephd,Benisty2009}.}
\end{figure}

The ABCD beam combination is an interferometric scheme\cite{labeyephd,Benisty2009} giving {\color{black}{a four point interferometric}} fringe sampling at its output.  {\elsaa{Each of the four outputs is the}} {\color{black}{combination}} of the two inputs with a different phase as represented in Fig.~\ref{fig:ABCDTheory}.
{\color{black}{ABCD cell}} input {\elsaa{1 and 2}}, i.e interfering beams, are split in two and are {\color{black}{recombined}} by pairs thanks to two directional couplers. {\elsaa{The phases of the four outputs}} are specified as follows: 

\begin{equation}
\begin{aligned}
\label{eq:equation8}
& A: \varphi_{12}^{A} \\
& B: \varphi_{12}^{B}=\varphi_{12}^{A}+\pi \\
& C: \varphi_{12}^{C}=\varphi_{12}^{A}+\pi/2\\
& D: \varphi_{12}^{D}=\varphi_{12}^{A}+\pi/2+\pi=\varphi_{12}^{A}+3\pi/2
\end{aligned}
\end{equation}

In order to build an achromatic {\color{black}{ABCD cell}}, a $\pi/2$ achromatic passive phase shifter is required. A phase shifter is composed of two parallel waveguides: {\color{black}{waveguide 1 and 2}}. Each {\color{black}{waveguide}} {\elsaa{of the same total length,}} is divided into N segments of optimized widths and lengths. {\elsaa{The width variations induce variations of the mode effective index, and thus a difference in the optical path length between the two arms}}. The phase difference for a phase shifter of length L composed of a N=1 segment is given as follows: 

\begin{equation}
\label{eq:equation9}
\Delta\varphi=\varphi_1-\varphi_2=\frac{2\pi}{\lambda}(n_{eff,1}-n_{eff,2})L
\end{equation}

with $n_{eff,1}=A_1+B_1\lambda+C_1\lambda^2$ and $n_{eff,2}=A_2+B_2\lambda+C_2\lambda^2$ the effective indexes of waveguides 1 and 2. 
This {\color{black}{N=1 segment}} phase shifter is achromatic if: 

\begin{equation}
\label{eq:equation10}
\forall \lambda,\quad \frac{\partial \varphi}{\partial \lambda}=-\frac{2\pi}{\lambda}(\frac{\Delta n}{\lambda}-\frac{\partial \Delta n}{\partial \lambda})L=0 
\end{equation}

with $\Delta n=n_{eff,1}-n_{eff,2}$ the effective index difference between waveguide 1 and 2. Following P. Labeye's PhD thesis\cite{labeyephd}, these equations can be generalized for {\color{black}{a design composed of N $\geq$ 1}} segments shifting the phase by $\varphi_0=\pi/2$:

\begin{equation}
\label{eq:equation11}
\varphi=\frac{2\pi}{\lambda}\sum_{i=1}^{N}\Delta n_i L_i=\varphi_0
\end{equation}

\begin{equation}
\label{eq:equation12}
\frac{\partial \varphi}{\partial \lambda}=-\frac{2\pi}{\lambda}\sum_{i=1}^{N}(\frac{\overline{A_i}}{\lambda}-\overline{C_i}\lambda)L_i=0
\end{equation}
with $\Delta n_i=n_{eff,1,i}-n_{eff,2,i}=(A_{1,i}-A_{2,i})+(B_{1,i}-B_{2,i})\lambda+(C_{1,i}-C_{2,i})\lambda^2=\overline{A_i}+\overline{B_i}\lambda+\overline{C_i}\lambda^2$ the effective index difference between waveguide 1 and 2 in segment $i$. Equation~\ref{eq:equation12} {\color{black}{is}} true for all wavelengths only if:

\begin{equation}
\label{eq:equation13}
\sum_{i=1}^{N}\overline{C_i}L_i=0 \quad \textrm{and} \quad \sum_{i=1}^{N}\overline{A_i}L_i=0.
\end{equation}

Consequently, one can derive that:

\begin{equation}
\label{eq:equation14}
\sum_{i=1}^{N}(\overline{B_i}L_i)=\frac{\varphi_0}{2\pi}
\end{equation}

is also a condition for Eq.~\ref{eq:equation11} to be valid {\color{black}{at}} all wavelengths. {\color{black}{Thus,}} it is possible to find sets of $(\overline{A_i}, \overline{B_i}, \overline{C_i})$ parameters that will produce an achromatic phase shift over a given bandwidth.

\subsubsection{Design and simulation of an achromatic $\pi/2$ passive phase shifter}
\label{subsec:abcd_phase_shifter}
\begin{figure} [h]
\begin{center}
\begin{tabular}{l}
\includegraphics[height=6 cm]{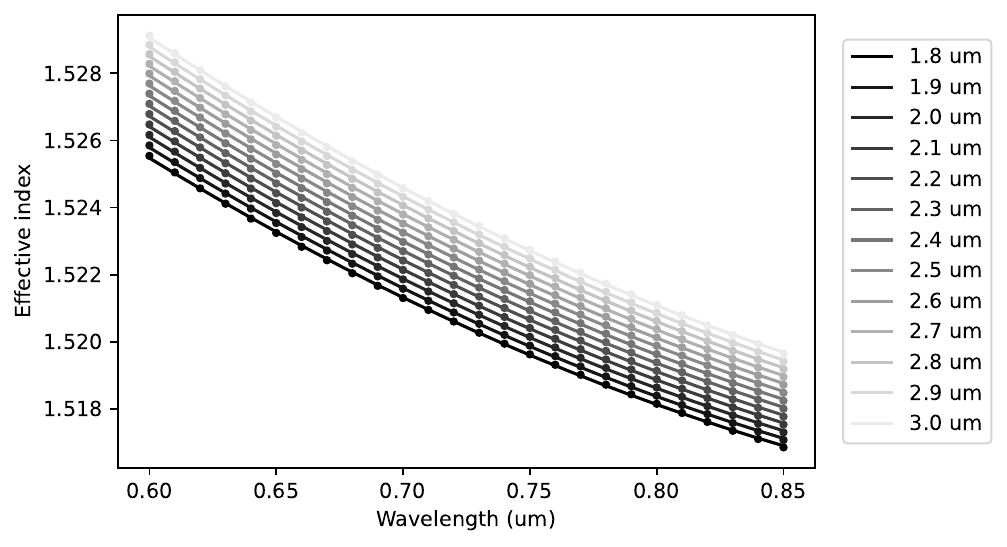}
\end{tabular}
\end{center}
\caption 
{\label{fig:ABCDNeff} 
{\elsaa{Effective indexes estimated with BeamPROP simulations (dots), as a function of wavelength and for different waveguide widths. The lines show the result of the second order polynomial fit.}}}
\end{figure}

For a given waveguide width, the coefficients $A$, $B$ and $C$ are estimated based on simulations performed with the BeamPROP software. Effective indexes are computed for widths between $1.8$ and $3$\,$\mu$m with a step of $0.1$\,$\mu$m, as shown in Fig.~\ref{fig:ABCDNeff}. A second order polynomial fit is used to estimate the $A$, $B$ and $C$ coefficients for each waveguide width value. Two- and three-segments solutions for achromatic $\varphi_0=\pi/2$ phase shifters are investigated. The lithographic fabrication error on waveguide width ranges from 0.2 to 0.5\,$\mu m$ and is considered homogeneous throughout the wafer, the glass substrate on which several PICs are manufactured. Therefore, the width error $\epsilon_w$ is similar for the two waveguides of the phase shifter. Fig.~\ref{fig:ABCDNeff} shows that $n_{eff,w1+\epsilon_w}-n_{eff,w2+\epsilon_w} = n_{eff,w1}-n_{eff,w2}$ meaning that the phase shift does not change with the fabrication error if this error is homogeneous throughout the wafer. 

The system of equations to be solved is given by Eq.~\ref{eq:equation13} and Eq.~\ref{eq:equation14}, which can be rewritten with a matrix formalism: 
\begin{equation}
\label{eq:System}
\left[ L \right]=\left[ M \right]^{-1}\cdot \left[ \phi \right]
\end{equation}

with:\\

$L=\begin{bmatrix}
L_1 \\
L_2 \\
\end{bmatrix}$, $ M=\begin{bmatrix}
\overline{A_1} & \overline{A_2}\\
\overline{B_1} & \overline{B_2}\\
\end{bmatrix}$ and $\left[ \phi \right]=\begin{bmatrix}
0\\
\frac{\varphi_0}{2\pi}\\
\end{bmatrix}$ for the two-segment solution, and\\

$L=\begin{bmatrix}
L_1 \\
L_2 \\
L_3 \\
\end{bmatrix}$, $M=\begin{bmatrix}
\overline{A_1} & \overline{A_2} & \overline{A_3}\\
\overline{B_1} & \overline{B_2} & \overline{B_3}\\
\overline{C_1} & \overline{C_2} & \overline{C_3}\\
\end{bmatrix}$ and $\left[ \phi \right]=\begin{bmatrix}
0\\
\frac{\varphi_0}{2\pi}\\
0\\
\end{bmatrix}$ for the three-segment solution.
\\

For the N=2 and N=3 segments phase shifters, four and six waveguide widths values {\elsaa{have to be optimized}} respectively. This is done {\elsaa{by scanning all parameter grids, and }} computing segments lengths for all the possible width combinations. 
{\elsaa{When constraining the waveguide width between $1.8$ and $2.3$\,$\mu$m, i.e. the single-mode range for wavelength from $600$ to $820$\,nm, optimal parameter sets lead to achromatic $\varphi_0=\pi/2$ phase shifters about a dozen millimeters long. However they can be made shorter if multi-modal waveguides are use, i.e. width larger than $2.3$\,$\mu$m. Indeed, with larger width, the effective index increases and the light gets slowed down more efficiently. In this case, optimal parameter sets can be found for shorter total lengths: $1.8$\,mm for the two-segment design, and $2.7$\,mm for the three-segment design.}}

\begin{figure} [h]
\begin{center}
\begin{tabular}{ll}
\includegraphics[height=5.5 cm]{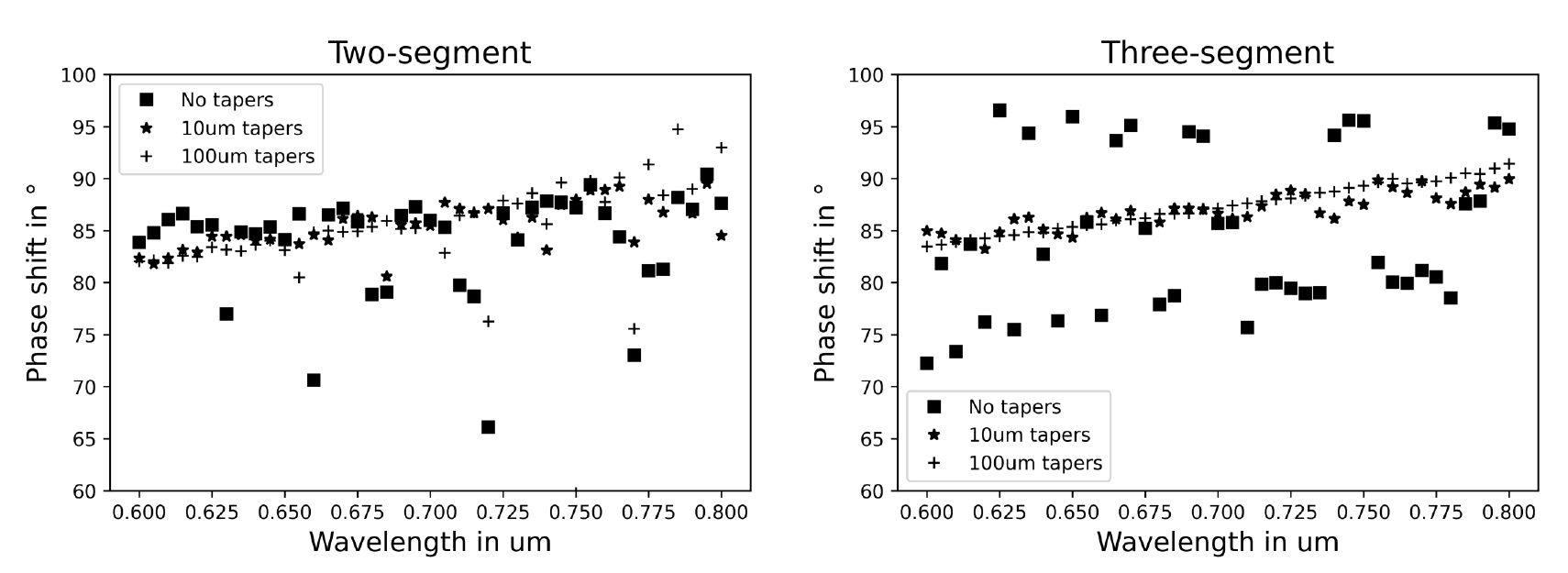}
\end{tabular}
\end{center}
\caption 
{\label{fig:ABCDSimu} Phase difference {\elsaa{achieved between the two arms of the optimized phase shifter}} as a function of wavelength {\elsaa{and for different taper options}}. \textbf{Left:} Two-segment solution {\color{black}{with a total length of $1.8$\,mm.}} \textbf{Right:} Three-segment solution {\color{black}{with a total length of $2.7$\,mm.}}}
\end{figure}

Mode coupling between segments is performed with tapers. Each taper present in one waveguide must be {\elsaa{included in the other one as well, as they would otherwise induce additional phase difference}}\cite{labeyephd}. Simulations are run in the three following cases: without tapers, 10\,$\mu$m and 100\,$\mu$m long tapers. Results are reported in Fig.~\ref{fig:ABCDSimu}, {\elsaa{highlighting the need for tapers and}} showing that an achromatic phase shift is achieved over the 600-800\,nm bandwidth {\elsaa{within an accuracy of 2 deg at best}}. The three-segment solution offers a better phase {\color{black}{shift}} performance at the cost of a longer total length as described in Table.~\ref{tab:tablePhaseshift}.

\begin{table}[h!]
\label{tab:tablePhaseshift}
\begin{center}       
\begin{tabular}{|c|c|c|}
\hline
\rule[-1ex]{0pt}{3.5ex} Number of segments & Tapers & Mean phase shift $\pm$ Standard deviation \\
\hline
\rule[-1ex]{0pt}{3.5ex}  & None & 84\,$^\circ$ $\pm$ 5\,$^\circ$ \\
2 & $10$\,$\mu m$ & 86\,$^\circ$ $\pm$ 2\,$^\circ$  \\
 & $100$\,$\mu m$ & 91\,$^\circ$ $\pm$ 41\,$^\circ$ \\
\hline
\rule[-1ex]{0pt}{3.5ex}  & None & 85\,$^\circ$ $\pm$ 8\,$^\circ$ \\
3 & $10$\,$\mu m$ & 87\,$^\circ$ $\pm$ 2\,$^\circ$  \\
 & $100$\,$\mu m$ & 87\,$^\circ$ $\pm$ 2\,$^\circ$ \\
\hline
\end{tabular}
\end{center}
\caption{\label{tab:tablePhaseshift} Simulated mean value and standard deviation of the phase shift for different optimized phase shifters. The mean value of the phase shift is computed over the $600$ to $800$\,$nm$ spectral band.}
\end{table}
\subsection{Latest prototypes and test functions}
\label{subsec:wafer}

\begin{figure} [h]
\begin{center}
\begin{tabular}{l}
\includegraphics[height=9 cm]{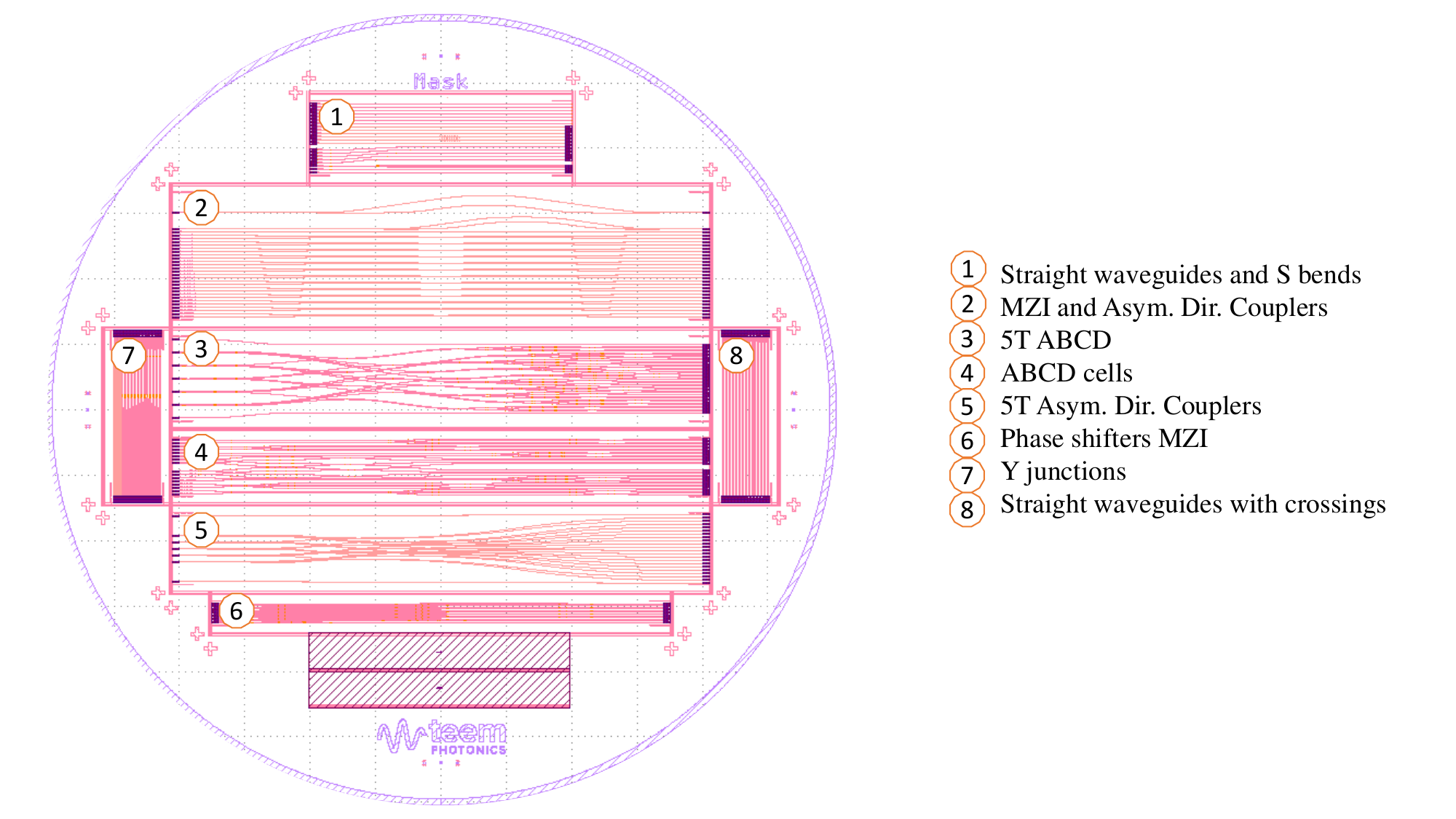}
\end{tabular}
\end{center}
\caption 
{\label{fig:Wafer} {\color{black}{Latest wafer Graphic Design System (GDS) view on Klayout software. {\elsa{It comprises test PICs with simple functions to characterize fundamental properties of the waveguides, as well as isolated combiner functions (asymetric directional couplers and ABCD cells), and also two complete 5T-combiners.}}}}}
\end{figure}

{\elsaa{A test wafer has been manufactured in order to characterize and validate the models and the designs presented in this paper, in particular regarding the asymmetric directional couplers and the ABCD cells. The laboratory characterization of the different components that have been included in this wafer will be the subject of a future communication.

The complete wafer layout is presented in Fig.\,\ref{fig:Wafer}. It comprises different test PICs, labelled from 1 to 8 in the figure. There are test PICs intended for characterizing the individual building blocks, as well as two complete 5T-combiners: 
\begin{itemize}
    \item PIC 1 is composed of straight and curved waveguides to assess some fundamental properties of the waveguides. The width of the straight waveguides range from 1.8 to 3\,$\mu$m with a step of 0.2\,$\mu$m in order to characterize the spectral range where the single-mode behavior is achieved. Various S bends with curvature radii ranging from 10 to 40\,mm will help determine the minimum acceptable curvature radius based on loss measurements. This PIC also contains straight waveguides with 10 crossings at various angles of 5, 10, 20, 30, and 45 degrees to evaluate excess losses due to waveguide crossing. PIC 8 contains additional straight waveguides with 10 or 20 crossings at angles ranging from 3 to 15 degrees.
    \item PIC 2 contains asymmetric directional couplers as presented in Sec.~\ref{subsec:asymx}, with various sets of parameters for the interaction zone geometry, i.e. the differential width $dw$ between the coupling waveguides and the interaction zone length $L$. Characterization will mainly consists in spectral and polarized transfer rate measurements. Because some fundamental values of TEEM Photonics technology are currently not well known, especially concerning birefringence, this PIC also contains Mach-Zehnder interferometers for spectral effective indexes measurement. Effective indexes in both polarizations will allow to refine the BeamPROP model and include both polarizations in simulations.
    \item PIC 4 contains 8 ABCD cells with various phase shifters. 
    \item PIC 6 contains Mach-Zehnder interferometers with phase shifters presented in Sec.~\ref{subsec:abcd_phase_shifter}. 
    \item PIC 7 contains Y junctions with different junction zone geometries.
    \item PIC 3 is a complete 5T interferometric combination PIC based on ABCD cells, while PIC 5 is a 5T PIC based on uniformly asymmetric directional couplers. Their design has been presented in Fig.\,\ref{fig:FIRSTv2BeamCombination}. They are meant to be installed on the FIRST instrument at the Subaru telescope, provided that their performance in terms of throughput and polarization is high enough.
\end{itemize}}}


\section{Conclusion}
\label{sec:ccl}


{\elsaa{Photonics integrated circuits constitute promising devices to perform beam manipulation requiring functions such as beam splitters, phase shifters or combiners. They thus offer stable and robust solutions for interferometric instruments, for long baseline interferometers, as well as pupil remapper instruments. While the technology has been well developed at the telecom wavelength range in the infrared, high performance visible PICs are still difficult to achieve. Within the SCExAO/FIRST project, a visible spectro-interferometer performing pupil remapping at the Subaru telescope, we are working with TEEM Photonics to produce a 5T beam combiner with high enough performance in terms of throughput and chromaticity, that would be suited for on-sky observations.

The development process is iterative and started with a first step to characterize the fundamental properties of the waveguides manufactured with the ioNext technology operated by TEEM Photonics. Based on the effective refractive index profile model of TEEM Photonics' standard waveguide validated by laboratory measurements, we have designed and optimized symmetric and asymmetric directional couplers, as well as achromatic phase shifters intended for ABCD cells, working in the 600-800\,nm wavelength range. A wafer containing various test PIC designs has recently been manufactured. The next step will be to characterize these PICs in the laboratory, to further evaluate losses due to propagation, bends or crossings, and to identify polarization dependent behaviors. These results will be the subject of a future paper.}}

\section{Acknowledgements}
A part of this work was previously published as a SPIE proceeding\cite{lallement2022} for the SPIE Astronomical Telescopes and Instrumentation 2022 event in Montréal, Québec, Canada. This project is supported by the French National Research Agency (ANR-21-CE31-0005) and the doctoral school Astronomy $\&$ Astrophysics of Ile de France (ED 127). {\color{black}{Authors also acknowledge  the funding by ASHRA (Action Spécifique Haute Resolution Angulaire) from INSU-CNRS and thank TEEM Photonics for their support and trust.}}
\bibliography{report}   
\bibliographystyle{spiejour}   
\vspace{2ex}\noindent\textbf{Manon Lallement (She/Her)} is a PhD student in instrumentation for astronomy at the Observatoire de Paris. Her PhD is supervised by Elsa Huby and Sylvestre Lacour and is funded by the doctoral school Astronomy \& Astrophysics of Ile de France (ED 127). She received her BS and MS degrees in photonics, theoretical and applied optics from the French Institut d'Optique Graduate School in 2019 and 2021, respectively.
Her current research interests include visible photonics for astronomical interferometry, 3D printed micro-lenses array for injection into single-mode fibers and interferometric data analysis. She is a member of SPIE. \\ \linkable{https://spie.org/profile/Manon.Lallement-4374172}

\vspace{1ex}
\noindent Biographies of the other authors are not available.

\end{spacing}
\end{document}